\documentclass[article, aps,pra,onecolumn,nofootinbib,superscriptaddress,longbibliography]{revtex4-2}
\UseRawInputEncoding
\usepackage[utf8]{inputenc}
\usepackage[normalem]{ulem} 
\usepackage[T1]{fontenc}
\usepackage[english]{babel}
\usepackage{amssymb}
\usepackage{amsmath}
\usepackage{amsfonts} 
\usepackage{amsthm}
\usepackage{mathtools}
\usepackage[ colorlinks=true,
  citecolor=blue,
  linkcolor=blue,
  urlcolor=blue
]{hyperref}
\usepackage{float}
\usepackage{enumitem}
\usepackage{bbold}
\usepackage{orcidlink}

\usepackage{threeparttable}
\usepackage{tikz}
\usepackage{siunitx,  
            booktabs, 
            caption}

\usepackage{ragged2e}
\AtBeginDocument{\justifying}

\newcommand{\greencheck}{}%
\DeclareRobustCommand{\greencheck}{%
  \tikz\fill[scale=0.4, color=green]
  (0,.35) -- (.25,0) -- (1,.7) -- (.25,.15) -- cycle;%
}

\newcommand{\orangecheck}{}%
\DeclareRobustCommand{\orangecheck}{
\textcolor{orange}{(} \kern-0.15cm
  \tikz\fill[scale=0.4, color=orange]
  (0,.35) -- (.25,0) -- (1,.7) -- (.25,.15) -- cycle;%
  \textcolor{orange}{)}
}

\newcommand{\orangeint}{}%
\DeclareRobustCommand{\orangeint}{%
  \tikz[baseline={([yshift=-.5ex]current bounding box.center)}]
  \node[scale=1.2pt, color=red]{?};
}

\begin{document}

\title{Time crystals on quantum devices}

\author{Gonzalo Camacho\orcidlink{0000-0001-6900-8850}}
\email{gonzalo.camacho@dlr.de}
\affiliation{Department High-Performance Computing, Institute of Software Technology, German Aerospace Center (DLR), 51147 Cologne, Germany}

\author{Benedikt Fauseweh\orcidlink{0000-0002-4861-7101}}
\email{benedikt.fauseweh@tu-dortmund.de}
\affiliation{Department High-Performance Computing, Institute of Software Technology, German Aerospace Center (DLR), 51147 Cologne, Germany}
\affiliation{Department of Physics, TU Dortmund University, Otto-Hahn-Str. 4, 44227, Dortmund, Germany}

\begin{abstract}
\justifying
Time crystals are nonequilibrium phases of matter characterized by the emergence of temporal ordering, in which an interacting many-body system develops robust structure in its time evolution that is not trivially dictated by the external driving or environment. While related phenomena have long been studied in classical nonlinear systems, their realization in entangled quantum matter represents a distinct frontier. The theoretical understanding of discrete time crystals has substantially advanced, yet recent experiments using modern quantum devices and quantum processors reveal regimes beyond established paradigms. These developments call for an extended classification of time-crystalline phases according to both their stabilization mechanisms and their physical character, including discrete and continuous, closed and open, critical, topological, quasiperiodic, and controlled realizations. We review recent implementations of time crystals on quantum platforms and propose such a classification framework, identifying promising directions for the discovery of novel time-crystalline phases of matter.
\end{abstract}

\maketitle 

\section{Introduction}\label{sec:intro}
The original idea of employing quantum devices as universal simulators~\cite{feynman_simulating_1982,lloyd_universal_1996} has motivated a rapid development in the field of quantum technologies in recent years, putting on a firm footing the capabilities brought by quantum devices to study nature at its most fundamental level. Even though currently available quantum devices are still susceptible to noise~\cite{preskill_quantum_2018}, impressive experimental achievements in the field of quantum simulation~\cite{bloch_quantum_2012,bloch_many-body_2008,blatt_quantum_2012,gross_quantum_2017} have opened up the possibility to explore interesting quantum phenomena employing a range of different platforms~\cite{houck_-chip_2012,aspuru-guzik_photonic_2012,bernien_probing_2017,joshi_probing_2022} where quantum advantage with respect to classical simulation is expected~\cite{cirac_goals_2012,daley_practical_2022}. The maturation of various different technologies for quantum devices has motivated further interest in the study of dynamics of many-body quantum systems~\cite{eisert_quantum_2015, fauseweh_quantum_2024}, in particular whether traditional concepts used in the definition of equilibrium phases of matter can be generalized to the out-of-equilibrium realm.

A nonequilibrium phase of matter that has received considerable attention in recent years is that of time crystals (TCs)~\cite{wilczek_quantum_2012}. While these systems have attracted an impressive amount of attention from a theoretical point of view, the possibility of implementing and observing time-crystalline behavior on current quantum devices has made these systems one of the most promising potential applications for the near-term use of quantum technologies.

\subsection{Aims and scope}\label{subsec:aims}

Time crystals have established themselves as an independent research field that has experienced a massive growth, receiving input from both theory and experiments. On the theory side, this has resulted in a number of reviews on the subject~\cite{sacha_time_2018,khemani_brief_2019,sacha_time_2020,else_discrete_2020,guo_condensed_2020,zaletel_colloquium_2023}, which focus mostly on discrete time crystals. However, the increasing diversity of time-crystalline phases, differing both in their physical character and in their stabilization mechanisms, indicates that a unified picture of time-crystallinity is still emerging. In this review, we therefore do not aim to provide an in-depth theoretical introduction to the topic. Instead, based on the recent advances and extensions of the concept, we provide a classification of the different types of time crystals that have been identified to date and provide an overview of their implementation on different quantum hardware. 

We focus on experimental realizations of time crystals that have been implemented exclusively on quantum devices, i.e., platforms that require a description by quantum mechanical laws. A genuine quantum mechanical time crystal requires that the system under study, and the device, contain a minimal, nonzero amount of entanglement. Including entanglement as a resource~\cite{horodecki_quantum_2009} establishes a direct connection with the fields of quantum information theory and quantum computing. Thus, we deliberately leave out cases concerning the simulation of classical systems by digitization of their equations of motion, because in this case, the system under study does not contain any entanglement and can in principle be simulated efficiently with a classical computer. At the same time, we include systems that are genuinely quantum mechanical in the sense of requiring entanglement for their physical description, even when their macroscopic dynamics admit an effective classical field theory or semiclassical description, thereby separating the notion of intrinsic quantumness from that of classical simulability.

In the literature, quantum entanglement acting as a necessary resource for realizing these phases is rarely explicitly discussed. Therefore, we aim to identify if the TC behavior, i.e., spontaneous time translational symmetry breaking, is likely the result of a quantum mechanism and entanglement potentially plays a role. As mentioned above this also includes systems where the dynamics can be  modeled by a semi-classical theory. In this case, the notion of entanglement and non-classical correlations can still be present.  A familiar example is provided by Bardeen-Cooper-Schrieffer (BCS) superconductors, where the superconducting state is a genuinely quantum many-body phase involving entanglement and non-classical correlations, such as those associated with spin-singlet pairing. Nevertheless, its essential physics can be captured by a mean-field description with effectively classical dynamics that can be simulated efficiently. 

To this end, we cover the entire timeline ranging from Wilczek's original 2012 proposal of a time crystal until the most recent experiments carried out in varied forms of quantum hardware. This time-span of fourteen years has resulted in a vast literature in the field, and thus we aim to provide a modular structure dividing the theoretical proposals from the experimental realizations.

The review is organized as follows. In Sec.~\ref{sec:theory}, we give a basic introduction to the theory of time crystals and their early proposals. We also provide a classification of time crystals based on their different stabilization mechanisms, providing an outline of both experimentally observed and outstanding cases. In Sec.~\ref{sec:quantum_devices}, we briefly present the different quantum platforms currently available for quantum simulation, addressing their individual advantages and limitations for implementing time crystals. Sec.~\ref{sec:implementations} reviews experimental results of time crystals on quantum devices, outlining current challenges based on their classification. In Sec.~\ref{sec:proposals}, we give a broad outlook on the outstanding theoretical proposals to realize time crystals on quantum devices, arguing which of these show stronger potential for immediate implementations. In Sec.~\ref{sec:conclusion}, we conclude the review.

\section{General theoretical concepts}\label{sec:theory}

\begin{figure}[!ht]
 \centering
        \includegraphics[width=0.8\textwidth]{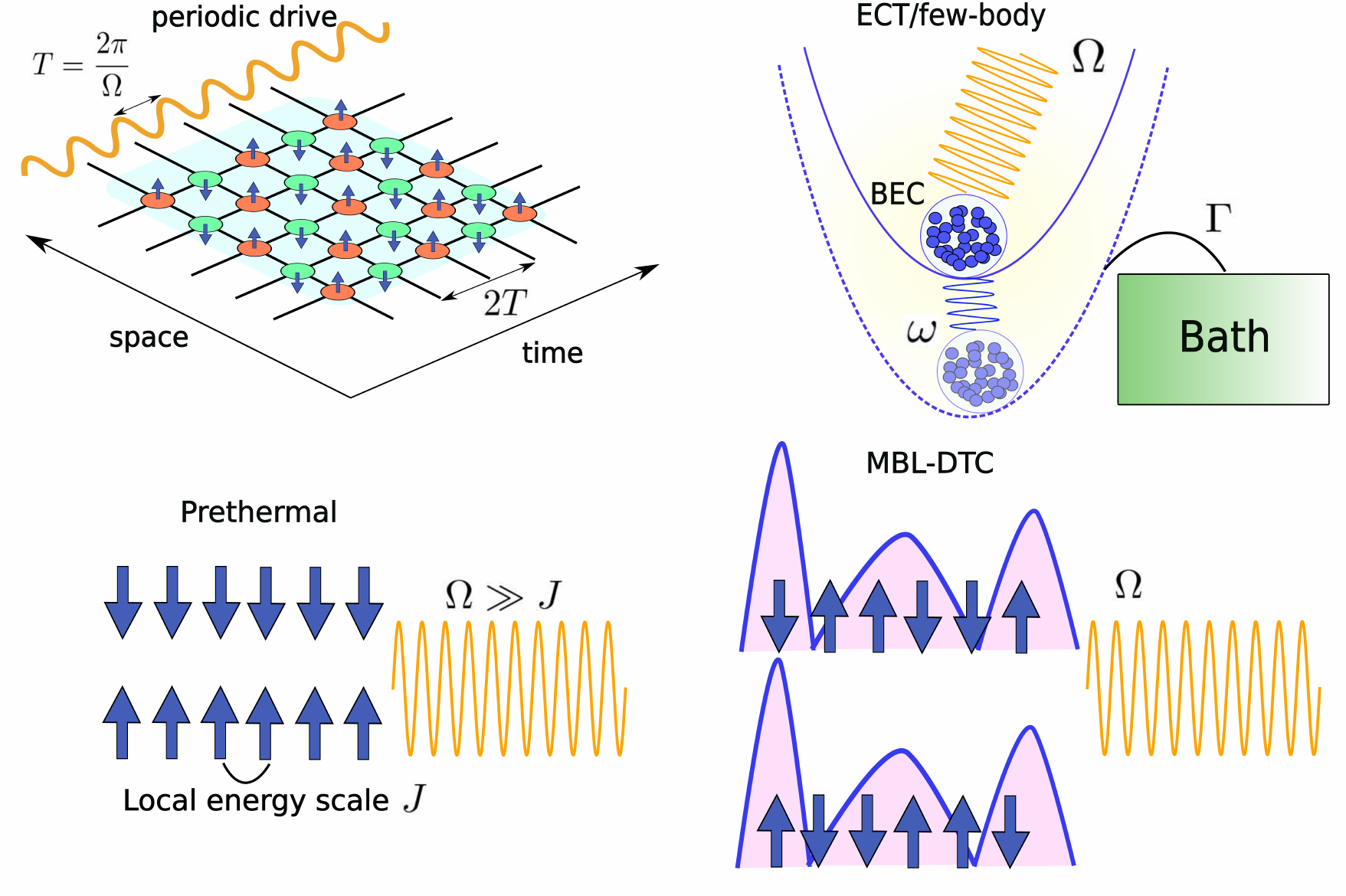}
 \caption{Schematic view of a DTC breaking the discrete symmetry of the drive. While the drive oscillates with period $T$, the system displays subharmonic motion with period $2T$. Time crystal phases can be stabilized through different mechanisms. 
 Prethermal DTCs require large frequencies $\Omega\gg J$ above some characteristic energy scale of the system $J$, usually from initially ordered (low temperature) states. In contrast, MBL-DTC is characterized by the system displaying subharmonic motion starting from random (high temperature) initial states, with the drive and the MBL phase coexisting. Some genuinely many-body quantum systems, such as Bose-Einstein condensates (BECs), display time crystal behavior under periodic driving and dissipation to a bath with some effective coupling $\Gamma$. In these cases, the stabilization mechanism is commonly described by effective classical theories (ECTs) or rather few-body descriptions involving mean-field approaches.}
\label{fig1}
\end{figure}

\subsection{Early works on spontaneous time translation symmetry breaking}\label{subsec:origins}
Time crystals were conceptualized as the time analog of ordinary spatial crystals. Roughly speaking, a time crystal is a macroscopic \emph{many-body} system in perpetual periodic motion with a different period from the one governing its equations of motion.~\footnote{Non-periodic systems can be regarded as having an infinite period of oscillation. A time crystal would break this infinite periodicity to oscillate at some finite rate $T<\infty$.} Among other properties, time crystals develop long-range order~\cite{sacha_time_2018,khemani_brief_2019,sacha_time_2020,zaletel_colloquium_2023,von_keyserlingk_absolute_2016,moessner_equilibration_2017} (LRO) along the temporal dimension of the space-time manifold, retaining infinitely long-lived memory about distant time events. The underlying mechanism of time crystal formation is that of spontaneous time translation symmetry breaking (S$\tau$B), where the state of the macroscopic system, be it classical or quantum~\cite{wilczek_quantum_2012,shapere_classical_2012,yao_classical_2020}, breaks any temporal symmetries governing its equations of motion. 

The idea of S$\tau$B was originally developed in the context of ground states of macroscopic systems~\cite{wilczek_quantum_2012}, with early proposals suggesting that time crystals could be experimentally observed under equilibrium conditions using a trapped-ion quantum device~\cite{li_space-time_2012}. Soon after, the possibility that S$\tau$B takes place under equilibrium conditions became a subject of debate~\cite{nozieres_time_2013,bruno_impossibility_2013,bruno_comment_2013,wilczek_wilczek_2013}, given that for isolated systems S$\tau$B challenges the second law of thermodynamics. These early objections were further supported by important no-go results showing that for local Hamiltonians no spatiotemporal LRO can emerge under equilibrium conditions~\cite{watanabe_absence_2015}, i.e., if a system experiences S$\tau$B, then the system must be out of equilibrium.

A generic example is a system that is driven by a $T$-periodic Hamiltonian, i.e., a system with a discrete time symmetry. The important distinction arises when the response of the system is observed. If a discrete symmetry is spontaneously broken resulting in a response correlated with the drive at a multiple of $T$, i.e., $nT,\, n \in \lbrace  2,3 \dots \rbrace $, then S$\tau$B takes place in a discrete fashion, and one talks about discrete time crystals (DTCs), see Fig.~\ref{fig1}. In frequency space, the system responds at a subharmonic frequency which is in clear correlation with the underlying periodicity of the drive. 

The DTC is the clearest manifestation of S$\tau$B under nonequilibrium conditions. In a different fashion, S$\tau$B can take place with the system presenting a somewhat uncorrelated response with the drive. This occurs mostly if the drive is characterized by some relevant energy scales and the system responds by oscillating with a frequency completely unrelated to those energy scales. Another relevant scenario occurs in continuously pumped systems, for instance when an amplitude-varying wave field is employed as a drive. In that case, the drive is no longer periodic due to the continuous amplitude ramping, but the system can eventually respond by oscillating with a well-defined frequency. 

These last two scenarios are related to the notion of a continuous time crystal (CTC). In the second of these, it is argued that a continuous time symmetry is broken; however, the definition of a continuous time symmetry in this case is somewhat challenging, as the system evolves under nonequilibrium conditions and generically possesses a complex dynamical structure without a continuous translational symmetry in time. The former scenario is somewhat easier to identify: if no correlation exists between the response of the system and the drive, we cannot identify the system as a DTC, even if the broken symmetries are discrete ones. These considerations hint at a more flexible time crystal classification beyond the current DTC/CTC dichotomy. 

In fact, this DTC/CTC distinction is rather limited from the purist point of view, especially regarding a truly universally accepted version of the CTC phase. As an example, it has been shown that in certain dissipative models, observation of a CTC in a rotating frame manifests as discrete or incommensurate time crystal behavior in the laboratory frame~\cite{cabot_nonequilibrium_2024}. Another viewpoint is that CTCs are directly associated with intrinsic energy scales $\omega_{\text{CTC}}$ of the system that can be activated from uncorrelated driving protocols~\cite{kongkhambut_observation_2024}, eventually leading to mode locking and DTC behavior if the frequency of the drive is adjusted to be an integer multiple of $\omega_{\text{CTC}}$. 

These viewpoints evidence that, although there is general consensus that DTCs are by now established time-crystal phases, the different forms in which CTCs can occur are much broader and their classification is by far more sparse. One fundamental aspect to understand this difference is that while DTCs can be realized in isolated quantum systems having time-periodic Hamiltonians, breaking the continuous time translation symmetry seems to require some degree of dissipation with the environment. Even though this has become the standard conceptual framework for emerging CTCs, this viewpoint will inevitably be tailored to the environment specifications, which introduces a considerable challenge towards developing a universal description of these phases. Nevertheless, recent theoretical progress in this direction identifies that the presence of certain antiunitary symmetries like parity-time (PT) in the master equation can lead to CTC behavior in a wide range of scenarios~\cite{nakanishi_2026}.

Another aspect of time crystals is that they never reach equilibrium because the system keeps finding itself indefinitely in repetitive periodic motion. Thus, for isolated systems, time crystals challenge the conventional understanding of equilibrium and thermalization inherited from statistical mechanics principles. For a time crystal, fundamental thermodynamic concepts like ergodicity and entropy are expected to display unconventional behavior compared to generic systems following the standard statistical mechanics paradigm.

Therefore, to understand the physics of time crystals, we need to address the causes leading to stabilization of time-crystallinity in macroscopic quantum systems. Moreover, we will see that there is a sharp distinction between time crystals occurring in the presence of an external environment (open) and those completely isolated from the environment (closed). The latter case is identified with time crystals that can be realized perfectly on ideal (noise-free) quantum computers, with the system undergoing purely unitary dynamics. For the former, some degree of dissipation is necessary to allow non-unitary evolution in the system in order to realize time crystal behavior.

\subsection{Ergodicity and thermalization in closed quantum systems}\label{subsec:ergodicity}

Consider an isolated quantum system prepared in some initial state $|\psi_0\rangle$ that belongs to a macroscopically large Hilbert space. The state evolves under a generic time-dependent Hamiltonian $H(t)$ from an initial time $t_0$ until a final time $t$ (using units with $\hbar=1$):  
\begin{eqnarray}\label{eq:psit}
    |\psi(t)\rangle=\hat{\mathcal{T}}e^{-i\int_{t_0}^t d\tau H(\tau)}|\psi_0\rangle,
\end{eqnarray}
where $\hat{\mathcal{T}}$ is the time-ordering operator. What is the expected long-time behavior of the system? 

In the absence of an extensive number of symmetries, an analogy with the ergodic hypothesis for classical systems can be drawn at the level of the Hilbert space: we expect for sufficiently long times that any state will be visited at equal rates, i.e., all states become equally likely~\cite{pilatowsky-cameo_complete_2023,pilatowsky-cameo_hilbert-space_2024}. Intuitively, this occurs due to the absence of energy conservation, interactions, and the impossibility to dissipate energy to any external environment. In the long-time limit, the system is driven towards an infinite temperature state, maximizing its entropy.~\footnote{Strictly speaking, this cannot occur globally under unitary evolution, because the state remains pure at all evolution times. Instead, any \emph{subregion} of the system is characterized by the reduced density matrix in that particular region, and in the long time limit this becomes effectively a thermal density matrix with parameter $\beta=\frac{1}{k_B T}\sim 0$. Thus, any local probe of the system in this subregion resembles that of a maximally mixed state due to continuous absorption from the drive.} This generic view in the quantum case establishes an analogy with the second law of thermodynamics of classical systems. 

A different but related situation occurs when the system evolves on its own under a time-independent Hamiltonian, eventually approaching equilibrium. If in the long-time limit the values of local observables coincide with those predicted by an appropriate statistical equilibrium ensemble, it is said that the system has \emph{thermalized}~\cite{polkovnikov_colloquium_2011,nandkishore_many-body_2015,mori_thermalization_2018}. If the initial configuration $|\psi_0\rangle$ is chosen to have support in the eigenbasis of $H$ in the interval $\Delta E=\left[E-\delta E,E+\delta E\right]$, from Eq.~\eqref{eq:psit} any long-time average of a local observable $\hat{O}$ yields:
\begin{eqnarray}
    \lim_{t\to\infty}\frac{1}{t}\int_{0}^t \mathrm{d}t' \langle\psi(t')|\hat{O}|\psi(t')\rangle\approx \sum_{\varepsilon\in\Delta E}|\langle \varepsilon|\psi_0\rangle|^2\langle\varepsilon|\hat{O}|\varepsilon\rangle.
\end{eqnarray}
Given that for a macroscopic system energy level separation scales as $\delta\varepsilon\sim\text{exp}(-kN)$ for some constant $k$, there are exponentially many contributions in the ensemble average and thus statistical fluctuations in the eigenstates are expected to be negligible~\cite{deutsch_eigenstate_2018}. This results in having equivalent contributions in the weights $|\langle \varepsilon|\psi_0\rangle|^2$, yielding the microcanonical ensemble average. The above is known as the Eigenstate Thermalization Hypothesis~\cite{deutsch_quantum_1991,srednicki_chaos_1994,rigol_thermalization_2008} (ETH). 

Systems following ETH are frequently identified as ergodic~\cite{rigol_alternatives_2012,dalessio_quantum_2016}. The ETH is not a universal description for the late time evolution of generic isolated quantum systems, but it is somewhat expected to provide a common ground for many cases. For generic interacting many-body systems without an extensive set of conserved quantities, the ETH is expected to hold when the underlying dynamics is quantum chaotic~\cite{deutsch_quantum_1991,srednicki_chaos_1994,dalessio_quantum_2016}. Quantum chaos is commonly diagnosed through the spectral properties of the Hamiltonian: after resolving all conventional symmetries, nonintegrable systems typically exhibit level statistics consistent with the appropriate random-matrix ensemble, most notably Wigner–Dyson level-spacing statistics~\cite{bohigas_characterization_1984,dalessio_quantum_2016}. Correspondingly, the onset of such spectral correlations in many-body systems is generally accompanied by the emergence of ETH behavior~\cite{rigol_thermalization_2008,rigol_alternatives_2012}. However, there are mechanisms by which an isolated quantum system might fail to display ergodic behavior, normally due to the existence of an extensive number of internal or emergent symmetries. Examples in this category concern integrable systems~\cite{calabrese_introduction_2016}, or non-interacting, disordered systems showing signatures of Anderson's localization~\cite{anderson_absence_1958,evers_anderson_2008}.

\subsection{Ergodicity breaking mechanisms in driven closed quantum systems}\label{subsec:closed_sys}

By definition, a time crystal cannot satisfy the ergodic hypothesis. Due to its infinitely long-lived periodic motion, the system does not populate all available microstates at equal rates, but only those configurations that repeat within a single periodic cycle. Thus, a mechanism is required to circumvent the ergodic hypothesis for a time crystal.

Generalizing the concept of off-diagonal long-range order to S$\tau$B, there are two basic defining requirements common to any definition of a time crystal: (i) the system presents S$\tau$B in the thermodynamic limit in some local order parameter, (ii) it is a many-body system described by a Hamiltonian containing sufficiently short-ranged interactions. Thus, in accordance with (ii), the following discussion applies to interacting systems.

It should be stressed that additional supporting conditions defining a many-body quantum time crystal have been put forward in detail in the literature~\cite{sacha_time_2018, khemani_brief_2019, sacha_time_2020,watanabe_absence_2015}. Proposals for time crystals that fall outside this paradigm have also gained considerable attention in recent years, broadening the requirements to include systems hosting long-range interactions~\cite{kozin_quantum_2019}, topological order~\cite{giergiel_topological_2019}, time quasi-crystals~\cite{else_discrete_2020} or even disordered but structured drives~\cite{zhao_temporal_2023}. All these cases have in common the presence of at least one ergodicity breaking mechanism leading to time crystal behavior. 

An important set of problems where different ergodicity breaking mechanisms have been clearly identified concerns periodically driven quantum systems and DTCs~\cite{else_floquet_2016,khemani_phase_2016}. 
Due to the periodic symmetry, the dynamics in these systems are governed by the single-period Floquet operator,
\begin{eqnarray}\label{eq:floquet_uf}
U_F(T,0)\coloneqq\mathcal{T} e^{-i\int_0^T d\tau H(\tau)} =e^{-iH_F T},
\end{eqnarray} 
where $H_F$ is the Floquet Hamiltonian, i.e., the Hermitian operator generating the unitary $U_F(T,0)$. Since the Hamiltonian is explicitly time-dependent, energy is not a conserved quantity in these systems. However, the eigenstates of $H_F$ can be associated with quasi-energies uniquely defined modulo $\frac{2\pi}{T}$, establishing a close connection with systems evolving under a stationary, effective Hamiltonian. Even for time-dependent Hamiltonians $H(t)$ hosting short-range interactions, $H_F$ is generically nontrivial and usually involves long-range interactions.

The long-time dynamics of generic periodically driven systems are well understood~\cite{dalessio_many-body_2013,dalessio_long-time_2014,lazarides_equilibrium_2014,ponte_periodically_2015} and follow the set of arguments exposed in Sec.~\ref{subsec:ergodicity} for ergodic behavior. Even though in this case the Hamiltonian of the system is time-dependent, the long-time behavior of the system shares analogies with the ETH previously discussed in static Hamiltonians. The main difference with respect to time-independent systems is that, for interacting and periodic Hamiltonians, the system continuously absorbs energy from the drive (Floquet heating), and any subregion of the system tends to maximize its entropy, eventually reaching an infinite temperature (maximally mixed) state. In contrast, for time-independent Hamiltonians satisfying the ETH, local probes of the system in the long-time limit do not generally correspond to such an infinite temperature state. To distinguish these two scenarios, we adopt the term Floquet ETH~\cite{mori_thermalization_2018} when referring to the generic fate of periodically driven systems. 

This generic view finds its own limitations in periodically driven systems, with certain models finding exceptions in the thermodynamic limit~\cite{prosen_time_1998, prosen_ergodic_1999,luitz_absence_2017,haldar_dynamical_2017,haldar_onset_2018,harper_topology_2020}. Remarkably, these models present short-ranged interactions in the Hamiltonian, and it is in general an open question how \emph{local} their effective Hamiltonian associated with the evolution operator in Eq.~\eqref{eq:floquet_uf} is. As of now, there does not seem to be a universally accepted mechanism for this rather unusual behavior of these periodic systems, where identifying all relevant symmetries, especially for the unitary evolution operator, constitutes a rather nontrivial task. For instance, while Ref.~\cite{prosen_time_1998} points towards the existence of intermediate regions interpolating between completely integrable and chaotic regimes in the spectrum statistics of the Floquet unitary operator, in Ref.~\cite{luitz_absence_2017} moderate interactions can result in transport suppression caused by interparticle collisions. In disordered driven systems, the presence of Floquet symmetry-protected topological (FSPT) phases can lead to eigenstate order~\cite{harper_topology_2020} preventing Floquet heating from rapidly building up in the system. Due to the presence of quasi-conserved quantities, some periodically driven systems can actually behave in a non-ergodic way in the long-time limit. In general terms, there are two main escape routes to avoid the ergodic fate: (i) Prethermalization taking place in systems driven at very high frequencies~\cite{bukov_universal_2015,kuwahara_floquetmagnus_2016,mori_rigorous_2016,weidinger_floquet_2017,else_prethermal_2017,abanin_exponentially_2015,abanin_effective_2017,abanin_rigorous_2017,mori_thermalization_2018,luitz_prethermalization_2020,machado_long-range_2020,ho_quantum_2023}; and (ii) periodically driven systems hosting strong spatial disorder~\cite{lazarides_fate_2015,ponte_periodically_2015,abanin_theory_2016,bordia_periodically_2017}. 

Prethermalization in periodically driven systems occurs when the driving frequency $\Omega=\frac{2\pi}{T}$ is very large compared to the intrinsic energy scales of the system~\cite{bukov_universal_2015,abanin_exponentially_2015}. In such cases, the dynamics are governed by a time-independent, quasi-local, effective Hamiltonian ($H_{\text{eff}}$)~\cite{goldman_periodically_2014} that approximates the true Floquet Hamiltonian $H_F$ up to exponentially long times, so that energy absorption in the system is suppressed~\cite{abanin_exponentially_2015} by the high drive frequency $\Omega$. As a result, after all transient behavior has died out, the dynamics of the system enters a prethermal phase with a slow heating rate that lasts exponentially long $t_*\sim \text{exp}(c\Omega)$, with $c$ a constant and $\Omega$ the driving frequency. During this prethermal time window, the system absorbs energy from the drive at a very small pace, which eventually leads in the long-time limit to the fate of Floquet ETH that we previously mentioned. 

Although the applicability of prethermalization is not limited to short-range interacting models~\cite{abanin_rigorous_2017, kuwahara_floquetmagnus_2016}, this mechanism has its own limitations, namely that any direct observation is very sensitive to the choice of initial conditions. In particular, in the absence of global $U(1)$ symmetries, only initial configurations that have finite support on edge states in the spectrum of $H_{\text{eff}}$ are good candidates for showing prethermal behavior. Additionally, employing very high-frequency drives is a necessary condition to achieve prethermalization, which often translates into employing drives that are orders of magnitude above the intrinsic energy scales of the system. 

An interesting aspect of prethermal phases is that they can host unconventional non-equilibrium states for times $t\leq t_*$. For example, if the effective Floquet Hamiltonian contains an emergent symmetry below a certain critical temperature $T_c$, a suitably chosen symmetry broken initial state corresponding to $T<T_c$ can result in S$\tau$B and the emergence of a prethermal DTC~\cite{else_prethermal_2017}. In this case, the existence of a quasi-conserved quantity in the Floquet evolution operator is what causes slow heating absorption and the presence of a prethermal phase. There is the possibility of realizing prethermal time crystal behavior without spontaneous symmetry breaking in the effective Floquet Hamiltonian, but rather through the emergence of a global, long-lived $U(1)$ symmetry in the system that approximately commutes with the effective Hamiltonian~\cite{luitz_prethermalization_2020}. In this scenario, not only low-temperature initial states display nontrivial dynamics under periodic driving, but also high-temperature initial states that explicitly break the $U(1)$ symmetry. In summary, prethermal time crystals present themselves as a transient escape route from ergodicity, but eventually Floquet ETH is expected to set in. In addition, they are sensitive to variations in the symmetries of the Floquet unitary Eq.~\eqref{eq:floquet_uf} and the initial state choice.

For disordered systems, the generalization of Anderson's localization phenomenon to interacting systems is known as Many-Body Localization (MBL)~\cite{gornyi_interacting_2005,basko_metalinsulator_2006,nandkishore_many-body_2015,abanin_colloquium_2019}, and is widely believed to be an ergodicity breaking mechanism in low-dimensional systems, including non-periodic ones. The MBL phase is characterized by an emergent, extensive number of conserved quantities (the so-called ``l-bits"~\cite{nandkishore_many-body_2015,abanin_recent_2017,abanin_colloquium_2019}). In simple terms, localization introduces an extensive number of quasi-local conserved quantities in the system, leading to an effective integrability that breaks ergodicity. Although rigorous results on MBL for one-dimensional spin chains exist~\cite{imbrie_many-body_2016,deroeck2025absencenormalheatconduction}, the question of the stability of the MBL phase in the presence of small ergodic patches (quantum avalanches) is still an ongoing topic of research~\cite{de_roeck_stability_2017,luitz_how_2017,morningstar_avalanches_2022}. Specifically, the stability of localization-protected quantum order~\cite{huse_localization-protected_2013} is still debated, especially in systems containing long-range interactions~\cite{nandkhisore_2017prx} or coupled to an external environment. Thus, while systems undergoing MBL and localization protected quantum order stand as strong candidates to escape the ergodic regime, their applicability domain is generally limited to addressing low-dimensional, isolated systems. 

In the presence of a periodic drive, systems hosting an MBL phase avoid rapid energy absorption in a different way than systems experiencing prethermalization. First, having a high-frequency drive is not required a priori. Rather, it becomes a competing effect between the driving frequency and the disorder strength leading to localization in the system. Contrary to the prethermal case, the number of initial configurations that potentially break the Floquet ETH in the long-time limit is substantially larger, precisely due to the extensive number of emergent quasi-symmetries. The question is then whether the Floquet Hamiltonian in Eq.~\eqref{eq:floquet_uf} undergoes MBL. If that is the case, any choice of an initially uncorrelated state is expected to retain memory of its initial conditions in the long-time limit and thus break the Floquet ETH. More importantly, if S$\tau$B takes place, a many-body localized discrete time crystal (MBL-DTC) emerges, with the system responding subharmonically to the discrete symmetry of the drive~\cite{ponte_many-body_2015,khemani_phase_2016,else_floquet_2016}. This form of ergodicity breaking mechanism emerges due to the presence of robust eigenstate order in the localized phase~\cite{huse_localization-protected_2013,von_keyserlingk_absolute_2016,von_keyserlingk_phase_2016,yao_discrete_2017}, making the MBL-DTC independent of the initial configuration.

Even though the above mechanisms are considered to be central for witnessing ergodicity breaking dynamics in periodically driven systems, exceptions exist. The first is that of observed time crystal behavior in disordered dipolar ensembles~\cite{choi_observation_2017}. One of the main features of these time crystals is that their spatiotemporal correlations decay algebraically, in contrast to MBL-DTCs (where correlations do not decay) or prethermal DTCs (where correlations decay exponentially slowly with driving frequency). Although these systems are highly disordered, sufficiently long-range interactions are expected to induce delocalization effects~\cite{yao_many-body_2014,burin_energy_2006}, which in principle rules out the existence of MBL in these systems. Prethermalization is also ruled out, since the initial state is not a low-energy eigenstate of the effective prethermal Hamiltonian. Instead, what occurs is that an interplay between disorder-induced localization tendencies and delocalization effects from long-range interactions takes place, giving rise to a \emph{critical} time crystal~\cite{ho_critical_2017} hosting power-law decaying correlations. 

The second exception has to do with driven systems whose underlying dynamics is based on effective classical theories (ECTs) or few-body descriptions. This situation is expected in time crystals emerging from periodically driven Bose-Einstein condensates~\cite{sacha_modeling_2015}. In these systems, dynamics can be generally understood by an effective classical description in the thermodynamic limit where the Kolmogorov-Arnold-Moser (KAM) theorem applies, or where the governing equation of motion becomes a periodically driven version of the Gross-Pitaevskii equation (GPE)~\cite{zaletel_colloquium_2023}. Alternatively, an effective low-energy theory of these systems admits a two-boson mode description~\cite{sacha_modeling_2015}. A curious fact is that S$\tau$B can occur when the number of particles tends to infinity while the effective number of degrees of freedom remains finite, as in mean-field or few-mode descriptions~\cite{wang_discrete_2021}, with numerical results supporting this claim~\cite{kuros_phase_2020}. The ECT/few-body description quite often relies on introducing large approximations like mean-field theory into the system. Although this might be convenient for an amenable treatment of certain problems, it generally leaves open the question of the true effect of interactions in the system dynamics. Recent results for certain time crystal types predict considerable effects on the physical response of the system when moving away from solutions obtained by mean-field~\cite{liu_boundary2026}.

The ergodicity breaking mechanisms identified here are central in time crystal stabilization, and can be compared to each other by addressing their most prominent limitations. For instance, while MBL and localization protected eigenstate order stand out as routes to short-ranged models of time crystals in disordered systems, possible extensions to higher dimensions or dissipative regimes seem challenging~\cite{droenner2019}. On the other hand, prethermal systems are better suited for witnessing time crystal behavior provided the appropriate symmetries and low-energy states are identified. However, these systems are generally less tunable and limited to high-frequency drives, low-energy initial states or the existence of global $U(1)$ symmetries. TCs undergoing ECT/few-body stabilization mechanisms provide a promising route to find a wide range of yet unexplored time crystal phases, at the cost of relying on strong approximations and, to some degree, semiclassical approaches. Even if S$\tau$B is present, we want to discriminate between cases with genuine many-body correlations and those where perturbative or few-body descriptions provide a sufficient explanation of the observed response.

\subsection{Stabilization mechanisms in open periodically driven quantum systems}\label{subsec:open_sys}

Contrary to the case of periodically driven systems described by unitary dynamics, the concept of ergodicity is not clearly defined in the presence of an external environment. The effect of the environment on the system dynamics is frequently modeled as stochastic noise, whose Markovian~\cite{fazio_many-body_2025} or non-Markovian~\cite{breuer_colloquium_2016,de_vega_dynamics_2017} nature depends on the problem at hand.

The quantum dynamics of a system coupled to a Markovian environment are generally described by the Lindblad equation, which assumes that the reduced dynamics of the system are memoryless, i.e., the future evolution depends only on the present state. This is the case we will limit ourselves to, given that most up-to-date time crystal models of open quantum systems have considered the Markovian case~\cite{riera-campeny_time_2020}, although time crystals in the context of non-Markovian noise have also been theoretically explored~\cite{carollo_exact_2022}. For periodically driven systems coupled to dissipative environments, contrary to the closed-system scenario, it is expected that Floquet heating does not pose a problem because the absorbed energy from the drive can leak into the environment~\cite{mori_floquet_2023}. This energy leak into the environment can lead to stable time-crystal behavior, through the analogy of a limit-cycle in the dynamics.

Indeed, even in the MBL-DTC, which is considered to be the most stable and robust closed time crystal, any coupling to an external environment will generally result in the absence of DTC behavior~\cite{lazarides_fate_2017}. This result is in accordance with arguments questioning the stability of the MBL phase in the presence of a dissipative environment, although different approaches to circumvent this problem have been proposed~\cite{fischer_dynamics_2016,medvedyeva_influence_2016,levi_robustness_2016,luschen_signatures_2017,vakulchyk_signatures_2018,lenarcic_critical_2020}.

An alternative route to witness time crystal behavior in open quantum systems is to turn to the high-frequency regime, which can host prethermalization and emergent discrete symmetries of the associated Floquet operator that remain robust in the presence of dissipation. In this regime, coupling to an environment can indeed enhance the time crystal response in the weak coupling limit~\cite{vu_dissipative_2023}, until stronger coupling to the environment supresses S$\tau$B. This competing effect scenario is reminiscent of the critical time crystals previously discussed, but now the competing mechanisms are the coupling to an environment and the presence of very large driving frequencies. The result is a DTC with exponentially decaying correlations, with lifetime depending non-monotonically on the coupling to the environment. There remains the question of whether there are steady states for which S$\tau$B is preserved indefinitely, as in the MBL-DTC for closed systems. 

In contrast to the case of closed quantum systems, the question reduces to whether the system equilibrates towards a steady state that explicitly presents signatures of S$\tau$B~\cite{gambetta_discrete_2019}. This is the leading mechanism behind the so-called boundary time crystals~\cite{iemini_boundary_2018} (BTCs), where in the thermodynamic limit, the system evolves towards a steady state where only a reduced portion of the Hilbert space presents S$\tau$B (breaking the \emph{continuous} symmetry of time translation in that region).

\subsection{Operational frameworks}\label{subsec:op_frame}

Subharmonic response alone, is not sufficient to establish a DTC. In a similar manner, an uncorrelated periodic response to an external pump is not sufficient to establish a CTC. An ideal time-crystalline phase should additionally display rigidity of the subharmonic frequency against perturbations of the drive, robustness over a finite region of parameter space and a lifetime that diverges in the thermodynamic limit. This distinction is particularly important for finite-size quantum processors, where period doubling can occur as a transient response without constituting spontaneous breaking of discrete time-translation symmetry but also for prethermal time crystals, which ultimately thermalize after an exponentially long time scale.

These properties provide an operational framework for comparing different TCs and understanding their fundamental differences. Establishing these properties is also an important prerequisite for assessing their potential use in future applications, including quantum technologies.

Rigidity refers to the persistence of the time crystal frequency under perturbations of the system parameters, so that the observed oscillations remain locked to the symmetry-broken frequency rather than occurring only at a fine-tuned point~\cite{yao_discrete_2017}. The theoretical understanding of this rigidity depends strongly on the underlying stabilization mechanism. Provided that a stable MBL phase exists, MBL-DTCs are stable against weak local perturbations of the Floquet drive, while in prethermal DTCs the subharmonic response persists within a parametrically long-lived prethermal regime~\cite{yao_discrete_2017,kyprianidis_observation_2021}. Rigidity is, however, not restricted to these many-body mechanisms. In few-body classical systems, stable subharmonic responses can be understood through a time-dependent extension of the KAM theorem, which guarantees the robustness of quasiperiodic trajectories against sufficiently small perturbations~\cite{zaletel_colloquium_2023}. From an operational perspective, rigidity becomes particularly relevant when the oscillations themselves constitute a resource, as recently explored for autonomous quantum clocks~\cite{viotti_quantum_2025}.

Another important operational aspect of time crystals is their sensitivity to external perturbations. This is particularly relevant for developing quantum sensing applications~\cite{moon_discrete_2024} and performing quantum metrology tasks~\cite{montenegro_quantum_2023}. Importantly, robustness and sensitivity need not be contradictory: a TC may be robust against perturbations of the drive while remaining highly susceptible to perturbations coupling to its order parameter. This principle has been demonstrated for prethermal DTCs, where the time-crystalline response is robust against drive imperfections and inhomogeneities, while resonant external fields can strongly modify the TC lifetime and provide a frequency-selective sensing signal~\cite{moon_discrete_2024}. Prethermal DTCs nevertheless retain an initial-state dependence, since time-crystalline order occurs only when the energy density of the initial state places the effective prethermal Hamiltonian within its symmetry-broken regime~\cite{kyprianidis_observation_2021}. For MBL-DTCs, robustness against weak local perturbations can in principle persist indefinitely in the thermodynamic limit. More generally, a genuine many-body TC should display increasingly well-defined time-crystalline order with increasing system size, while the scaling of its lifetime reflects the underlying stabilization mechanism. 

\section{Quantum platforms}\label{sec:quantum_devices}

Here we give a summary of the current platforms employed for realizing quantum time crystals. Many of these platforms also serve as computing platforms that rely on digital, analog, or hybrid approaches. However, it should be noted that not all of them are suitable to perform computation tasks, e.g., dissipative BECs are limited in this regard.


Quantum devices based on superconducting (SC) qubits created from Josephson junctions have become increasingly active in recent years, promoted in part by their fast industrial development~\cite{blais_cavity_2004,clarke_superconducting_2008,krantz_quantum_2019,arute_quantum_2019,kjaergaard_superconducting_2020,google_quantum_ai_and_collaborators_quantum_2025}. Their favorable scalability to a large number of qubits and fast measurement rates have put SC qubits at the vanguard of research in the field of quantum simulation, even though qubit coherence times and overall noise reduction of these devices are still an ongoing topic of research. 

Trapped ions were originally proposed as one of the first quantum platforms~\cite{cirac_quantum_1995} to perform quantum computation~\cite{monroe_demonstration_1995} and simulation~\cite{blatt_quantum_2012,smith_many-body_2016,foss-feig_progress_2025}. Since then, these devices have seen remarkable progress in their development~\cite{ospelkaus_microwave_2011,lanyon_universal_2011,bruzewicz_trapped-ion_2019,monroe_programmable_2021}. One of the main characteristics of these platforms is that they present long coherence times resulting from controlled isolation of the qubits. Ions remain confined using strong electromagnetic fields (Paul traps), and provide a versatile platform to achieve arbitrary couplings between the qubits through controlled laser or microwave pulses addressing different energy levels. Current research on trapped-ion devices focuses on reducing measurement times and on finding efficient scalable approaches.

Neutral atoms~\cite{henriet_quantum_2020,graham_multi-qubit_2022,bluvstein_quantum_2022,semeghini_probing_2021} are also very promising platforms for the simulation of quantum time crystals. In these systems, the atoms are laser-cooled and their levels are addressed in order to encode qubits, which are entangled with one another using Rydberg states. These platforms offer promising scalability, flexibility in how qubits are locally addressed or non-locally coupled, and potential to transport entangled qubits in arbitrary geometries. 

Solid-state devices based on nitrogen-vacancy (NV) centers were experimentally realized as quantum registers~\cite{dutt_quantum_2007,maze_nanoscale_2008,abobeih_atomic-scale_2019}, and are now being actively used for quantum simulation~\cite{bradley_ten-qubit_2019,kucsko_critical_2018}. In these devices, a $^{14}$N vacancy sits between $^{13}$C diamond cells, with devices operating at room temperature. During single-qubit operations, the electron spin on the NV is optically excited, which in turn induces a hyperfine level splitting in one proximal nuclear spin.

Photonic quantum computing~\cite{knill_scheme_2001,flamini_photonic_2019} platforms are currently in development~\cite{barz_quantum_2015}. The resilience of photons against external noise makes them ideal candidates for the processing and transmission of quantum information employing linear optics elements. However, these devices suffer from scalability issues, as well as from difficulties in generating entangled operations employing photons. This can be circumvented by coupling the photons strongly to matter using resonators and Kerr non-linear optical microcavities~\cite{herr_temporal_2014,kippenberg_dissipative_2018}.

Alternative platforms are also being actively developed and explored for quantum simulation and for realizations of time crystals. Examples include quantum dots on semiconductor structures~\cite{yang_silicon_2019,west_gate-based_2019,wang_experimental_2022}, electron–nuclear spin systems in semiconductors~\cite{greilich_robust_2024} and cold-atom platforms coupled to a radiation field~\cite{ritsch_cold_2013}, as well as platforms of BECs~\cite{cornell_nobel_2002,ketterle_nobel_2002} and magnon condensates in superfluid helium~\cite{makinen_magnon_2024}. These have become highly relevant to realize a variety of time crystals, especially in the presence of dissipation. 

Another platform recently employed for realizing time crystal behavior is based on spin-maser devices~\cite{chupp1994,SATO2018588}, where self-driven oscillatory systems employ feedback schemes to compensate losses due to decoherence, thus maintaining the overall periodic motion over large atomic ensembles.

Recent years have also seen an impressive advance in the development of platforms capable of performing universal quantum computation beyond qubits by exploiting higher-dimensional quantum systems, whose basic unit of information is the \emph{qudit}. These platforms comprise a varied number of different physical setups, including trapped ions~\cite{ringbauer_universal_2022,hrmo_native_2023}, superconducting circuits~\cite{fedorov_implementation_2012,brock_quantum_2025,tripathi_qudit_2025}, ultracold molecules~\cite{vilas_optical_2024}, photonic circuits~\cite{kues_-chip_2017}, solid-state bosonic ladders~\cite{nguyen_empowering_2024}, silicon-photonic integrated circuits~\cite{chi_programmable_2022}, NV centers~\cite{soltamov_excitation_2019} and semiconductor-based platforms~\cite{moro_realization_2019,fernandez_de_fuentes_navigating_2024}.

\section{Implementations}\label{sec:implementations}

In this section we present an overview of the time crystals that have been observed on quantum devices. In particular, we provide a classification of these implementations according to the different stabilization mechanisms for the observation of temporal order. We also identify unobserved cases that show potential for implementation on available hardware.

\subsection{Many-body localized time crystals}

\begin{figure}[!ht]
 \centering
        \includegraphics[width=0.8\textwidth]{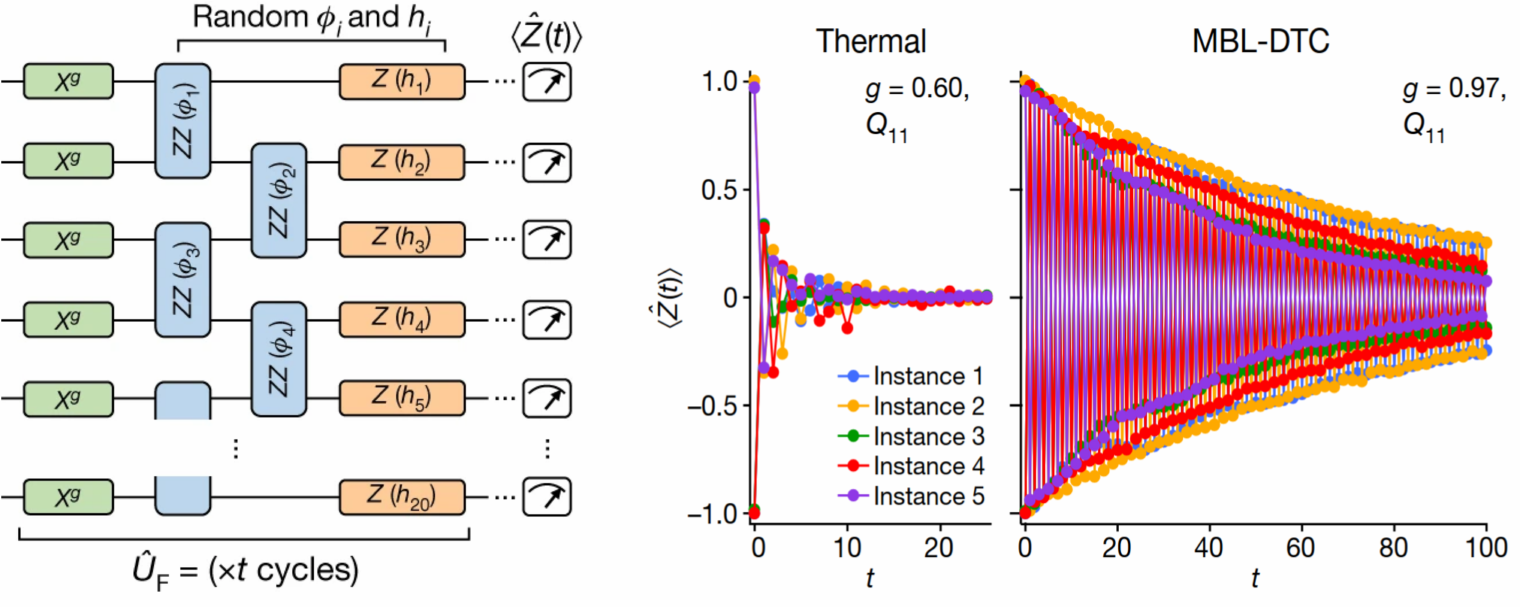}
 \caption{(Adapted from Ref.~\cite{mi_time-crystalline_2022}) Digital (gate-based) implementation of the MBL-DTC on a superconducting quantum processor, where the model consists of a periodically kicked 1D Ising chain with nearest-neighbor interactions and random local fields. In the MBL-DTC phase, different random instances of initial product states lead to robust subharmonic oscillations in the autocorrelation of all local observables, in contrast to the thermal phase where an arbitrary initial state choice leads to fast thermalization and absence of spatiotemporal order.}
\label{fig2}
\end{figure}

The first experiments observing time crystals under periodic driving were performed on trapped-ion devices; in particular, the MBL-DTC was reported to be observed in Ref.~\cite{zhang_observation_2017} on a trapped-ion device with all-to-all connectivity and power-law decaying interactions. Importantly, Ref.~\cite{zhang_observation_2017} states that the system undergoes MBL despite interactions being long-range with couplings decaying as a power law with exponent $\alpha>1$. The considered model is a periodically kicked Ising Hamiltonian, where a kick parameter serves to tune away from the perfect spin-flip scenario. The experiment reported evidence of robust subharmonic response in the total spin magnetization, identifying the phase diagram of the MBL-DTC by tuning the interaction strength and the kick parameter. The height of the peaks developed in the Fourier spectrum of the magnetization serves as an order parameter to identify the TC phase. However, due to the long-range nature of interactions and the highly polarized states, it is not clear that the true stabilization mechanism in this case is due to MBL, but rather points towards a prethermal time crystal~\cite{khemani_comment_2021}. The moderate system size reported in the experiment ($N=10$ spins) is far from faithfully mimicking results belonging to the thermodynamic limit.

Soon after, it was theoretically proposed that superconducting qubit-based platforms also represent suitable hardware candidates for the implementation of several MBL-DTC models~\cite{ippoliti_many-body_2021}. This analysis led to the first observation of the MBL-DTC in these platforms in Refs.~\cite{xu_realizing_2021,mi_time-crystalline_2022,frey_realization_2022}, where the model under consideration is a disordered, periodically kicked Ising chain. Refs.~\cite{mi_time-crystalline_2022,frey_realization_2022} witnessed subharmonic response in the local autocorrelation function of all qubits, which manifests pronounced peaks in their Fourier spectrum in the TC phase, signaling the emergence of spatiotemporal order. Additionally, Ref.~\cite{mi_time-crystalline_2022} reported results on an emergent spin-glass order parameter, supporting the notion that the MBL-DTC phase becomes a highly spatially correlated regime. Although these devices suffer from strong decoherence effects, clear signatures of subharmonic response could be witnessed for random initial ``bitstring'' states, a characteristic signature of the MBL-DTC, see Fig.~\ref{fig2}. Refs.~\cite{mi_time-crystalline_2022,frey_realization_2022} reached significantly larger system sizes ($N=20$ and $N=57$, respectively) than Ref.~\cite{zhang_observation_2017} supporting the survival of the MBL phase and robust subharmonic response in the limit of large system sizes.

Follow-up works on these platforms investigated the observation of MBL-DTCs further. In particular, Ref.~\cite{switzer_realization_2025} has reported experimental results on a superconducting qubit platform for a two-dimensional disordered version of the anisotropic Heisenberg model with signatures of discrete S$\tau$B in local autocorrelators and spin-spin correlations of the model. In addition, a genuine quantum information theoretic measure sensitive to multipartite entanglement is reported, the Quantum Fisher Information (QFI). 

A different category concerns topologically ordered time crystals in disordered ensembles. Digital quantum simulation of FSPT phases employing an array of 26 superconducting qubits~\cite{zhang_digital_2022} has reported S$\tau$B at the boundaries of the one-dimensional system, in stark contrast to the behavior in the bulk where no oscillations are present. The system is disordered and thus it is claimed that deep in the FSPT phase, MBL is present. However, both the presence of topologically protected edge modes and the three-body nature of interactions in the system could indicate that such a stabilization mechanism indeed goes beyond the presence of MBL. 

More recently, making use of superconducting qubit-based machines with the largest available number of qubits, critical properties of the MBL-DTC~\cite{hirasaki_shift_2025} have been investigated. For a 70-qubit superconducting quantum processor, subharmonic oscillations up to 25 cycles in the total magnetization were reported in the one-dimensional kicked Ising chain. Additionally, the variance of the order parameter as a function of the kick parameter showed a pronounced peak that remains robust with increasing system sizes even under strong decoherence effects. This system has also been used to address the performance characterization of these machines~\cite{zhang_characterizing_2023}.

The MBL-DTC has also been observed on a programmable spin-based quantum simulator comprising 27 $^{13}$C nuclear spins with all-to-all connectivity close to an NV center~\cite{randall_many-bodylocalized_2021}. In this case, an effective chain of dipole-dipole interacting spins is periodically kicked, where disorder is inherited from random positional placements of the nuclei in the system. An $N=9$ one-dimensional short-ranged model is effectively engineered to host the MBL phase. While the system size is much more limited compared to superconducting processors, the setup allowed up to $800$ DTC cycles to be observed for the correlation function.

For MBL-DTC realizations, superconducting qubits and trapped-ion implementations are the main drivers compared to alternative platforms. In the case of superconducting qubits, the main reason is the high degree of controllability of disorder strength and dimensionality offered by these platforms, alongside the ability to reach larger system sizes. In addition, trapped ions allow for long-range interactions to probe the destabilization of the MBL phase. In NV centers, where the number of qubits is low, the DTC could be observed for much longer as the qubits are extraordinarily well isolated from the environment.

\subsection{Prethermal time crystals}

\begin{figure}[t]
 \centering
        \includegraphics[width=0.9\textwidth]{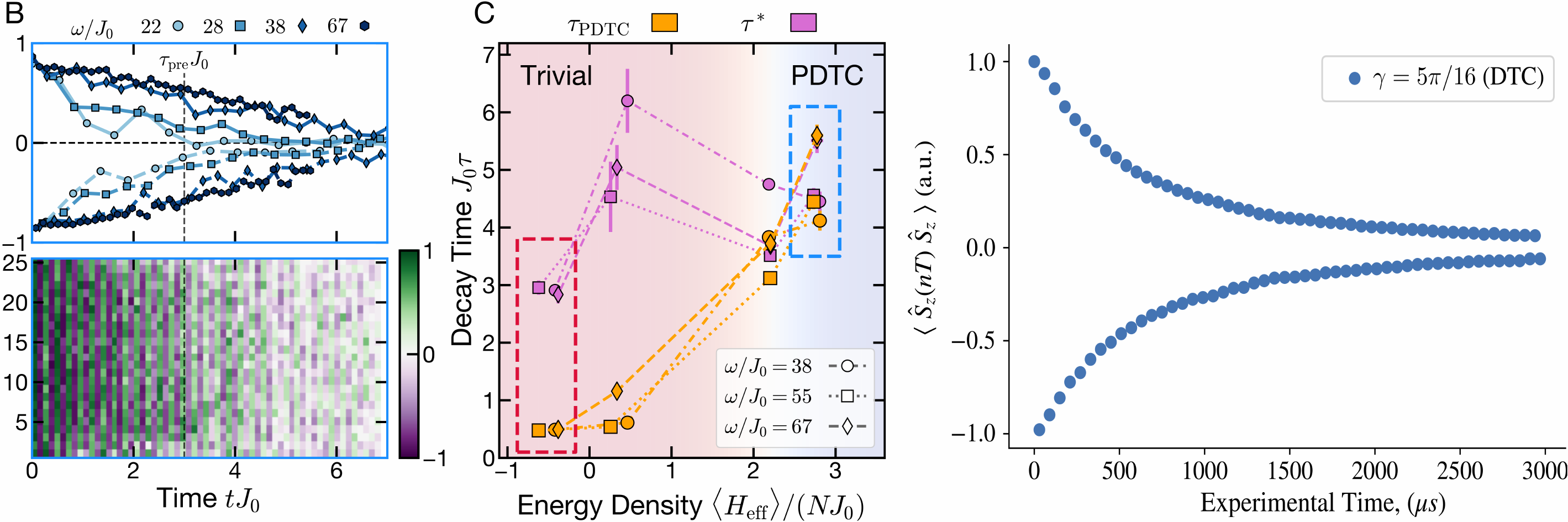}
 \caption{(Adapted from~\cite{kyprianidis_observation_2021} and~\cite{stasiuk_observation_2023}) Implementation of prethermal DTCs on a trapped-ion device~\cite{kyprianidis_observation_2021} (left) and the NMR quantum emulator~\cite{stasiuk_observation_2023} employing a fluorapatite ($\text{Ca}_5(\text{PO}_4)_3 \text{F}$) crystalline structure (right), with both systems hosting long-range interactions. In the experiment in Ref.~\cite{kyprianidis_observation_2021}, a spin-polarized initial state displays the slow decay in the magnetization dynamics (panel B). The survival of the prethermal phase up to finite thermalization time scales starting from low-energy density states of the effective Hamiltonian $H_{\text{eff}}$ is a distinct signature of prethermal DTCs (panel C). In the nuclear magnetic resonance (NMR) experiment of Ref.~\cite{stasiuk_observation_2023}, the initial state is represented by a classical (unnormalized) state of the form $\hat{M}_z=\frac{1}{2}\sum_l \sigma_l^z$. The system is kicked with $\gamma$ acting as a tuning parameter related to changing the period of the drive or the effective Hamiltonian interactions. Deep in the DTC phase, a clear subharmonic response of the autocorrelation function at stroboscopic times is reported. }
\label{fig3}
\end{figure}

The prethermal DTC was observed for the first time in Ref.~\cite{kyprianidis_observation_2021}, where the experiment reported observation of prethermal time-crystal behavior on a 25-ion trapped-ion quantum simulator, whose effective Hamiltonian is an all-to-all Ising Hamiltonian with uniform local fields, see Fig.~\ref{fig3}. The observation of subharmonic response in the local spin autocorrelations at high enough driving frequencies from polarized initial states is the characteristic signature of the prethermal discrete time crystal. In addition, the prethermal lifetimes are extracted directly from mean energy density decay curves, which helps characterize the frequency dependence of the Floquet heating time dependence and the crossover to the thermal phase. 

Prethermal DTCs have been observed to emerge and stabilize in the presence of long-range interactions on superconducting platforms~\cite{solfanelli_stabilization_2024}. While the magnetization shows a slow decay, the number of cycles restricted to $n\sim 10$ is too limited to witness the existence of a prethermal region in the dynamics. Moving away from purely digital implementations, Ref.~\cite{ying_floquet_2022} reported the observation of a $U(1)$ prethermal DTC under a digital-analog scheme, even in the presence of disorder.

The observation of a prethermal, topologically ordered DTC was reported in Ref.~\cite{xiang_long-lived_2024} on an 18-qubit quantum processor operating at periodic drives with $T\sim 1.4\, \mu$s of the corresponding circuit runtime, which ensures operation in the high-frequency regime where local magnetizations are shown to quickly decay without oscillating, and the presence of a DTC in this case is witnessed by subharmonic oscillations in rather quasi-local operators. 

There have been experimental realizations of prethermal DTCs on alternative platforms apart from trapped ions and superconducting qubits. For instance, in Ref.~\cite{beatrez_critical_2023}, a prethermal DTC was observed on a 3D system of $^{13}$C nuclei in diamond at room temperature. The system is driven in a way such that the carbon nuclei become hyperpolarized by surrounding NV centers, whose dynamics alternate between fast and slow drives. The process generates an effective quasi-conservation law, responsible for prethermalization. The observed lifetimes in the experiment are within the scale of seconds and up to $396$ Floquet cycles. Given the long-range nature of interactions, and the emergence of S$\tau$B starting from different initial state choices, it can be conjectured that the time-crystalline order observed in this case corresponds to a critical DTC.

In Ref.~\cite{moon_discrete_2024}, a prethermal DTC realized on a quantum processor of NV centers on $^{13}$C was employed to devise a frequency-selective quantum sensor of time-varying AC fields. The prethermal lifetime of the DTC is extended by orders of magnitude upon external action, a characteristic trait of controlled time crystals.

Solid-state NMR devices have also been employed to witness prethermal DTCs. In Ref.~\cite{stasiuk_observation_2023}, a crystal structure of Ca$_5$(PO$_4$)$_3$F was employed to obtain a quasi-1D nuclear spin ensemble, which is subjected to high-frequency periodic driving and displays subharmonic response of the spin-spin correlations and local autocorrelation functions. These measures are a genuine manifestation of long-ranged spatiotemporal spin correlations undergoing S$\tau$B in the model, which survive for lifetimes of the order of milliseconds. The crossover between prethermal and thermal phases is demonstrated as a function of the kicking parameter and interaction strength.

Going beyond qubit-based architectures, prethermal DTC behavior has been recently observed on a trapped-ion qudit processor~\cite{camacho_observing_2024}. In this case, a periodically driven $S=1$ chain is shown to undergo prethermal behavior due to a Hilbert-space localization mechanism. The emergent symmetry of the effective Hamiltonian locks initial low-temperature states into a symmetry sector of the Hilbert space. 

Recent advances in the emergent subfield of time rondeau crystals have led to experimental realization of this new form of prethermal spatiotemporal order on NV center quantum platforms~\cite{moon_experimental_2025}. 

Similar to the MBL case, the different platforms reveal a trade-off between programmability and stability of the prethermal response. Trapped-ion and superconducting processors offer high control over initial states and Floquet protocols, enabling direct tests of the dependence on initial energy density, the realization of different lattice geometries and symmetries, and higher-order subharmonic responses. Their accessible lifetimes, however, can be limited by decoherence and, for digital implementations of long-range interactions, by circuit depth and accumulated gate errors. Nuclear-spin platforms provide less microscopic control, but naturally occurring long-range interactions and long coherence times enable particularly rigid and long-lived responses, reaching hundreds of Floquet cycles in diamond. Recent superconducting experiments sustaining prethermal order beyond one hundred cycles nevertheless considerably narrow this gap. Thus, programmable processors are particularly advantageous for engineering and systematically probing prethermal phases, whereas spin ensembles provide a natural setting for accessing intrinsically stable, long-lived prethermal dynamics.

\subsection{Effective classical theories and few-body time crystals}

\begin{figure}[t]
 \centering
        \includegraphics[width=0.9\textwidth]{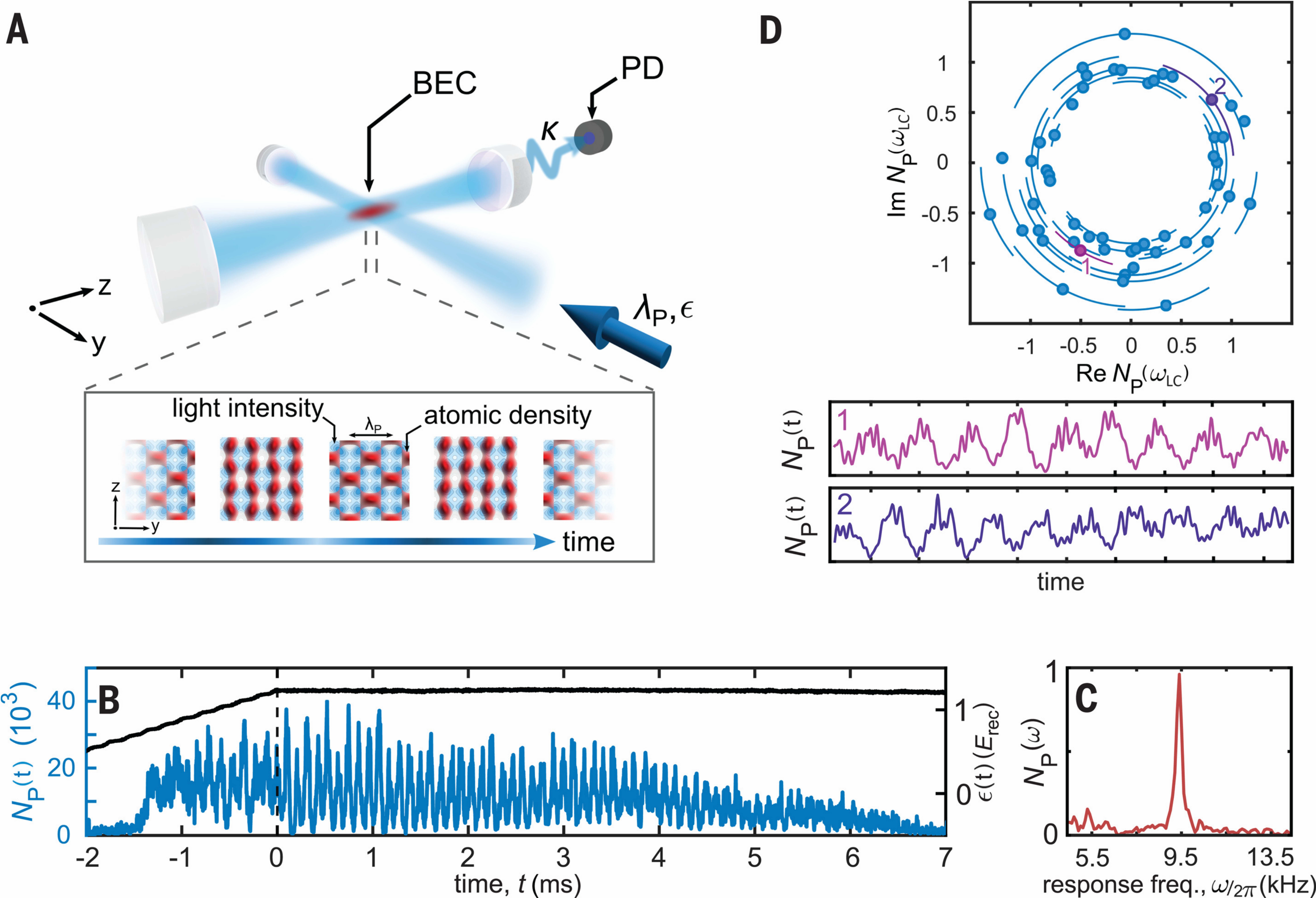}
 \caption{(Adapted from Ref.~\cite{kongkhambut_observation_2022}) Observation of a CTC employing a BEC in an optical cavity~\cite{kongkhambut_observation_2022}. In the experiment, the BEC is transversely driven by a highly polarized optical pump with increasing amplitude. After some pumping time, the amplitude of the drive is kept constant, and the system features subharmonic oscillations on the cavity photon number $N_P(t)$, which acts as the order parameter.}
\label{fig4}
\end{figure}

The first realization of a dissipative continuous time crystal (CTC) employed a Bose-Einstein condensate of $^{87}$Rb atoms in optical cavities in the presence of dissipation~\cite{kongkhambut_observation_2022}, see Fig.~\ref{fig4}. In this case, the intracavity photon number is the relevant order parameter for witnessing S$\tau$B as the optical pump strength of the laser is linearly ramped, and displays a clearly oscillating pattern, with lifetimes of a few milliseconds. The system sizes addressed in these experiments involve a large number of atoms ($\sim 10^{4}$), while the theoretical model describes an effective atom-field equation that incorporates nonlinear effects. In Ref.~\cite{kongkhambut_observation_2024}, a transition from the dissipative CTC to a dissipative DTC was experimentally observed employing a ramped pump protocol. 

Apart from CTCs, dissipative DTC behavior has also been reported in a BEC obtained from sodium atoms~\cite{smits_long-term_2020}, where oscillations of the radial breathing mode, which acts as the drive, are reported to have a different frequency from those of the observed crystalline mode, thus signaling S$\tau$B. The achieved lifetimes are on the order of seconds, with results showing clear stabilization when multiple modes are added to the system dynamics. In Ref.~\cite{huang_observation_2025} a CTC in noble spin gases of $^{129}$Xe and $^{87}$Rb atoms has been implemented through optical pumping. This experiment is at the boundary between open CTC and, as a result of the feedback, controlled TCs. In this case, the stabilization mechanism is few-body as the collective dynamics can be described by a set of nonlinear Bloch equations. A marked signature of time-crystal stability is the existence of pronounced peaks in the Fourier spectrum of the probe beam intensity even in the presence of noise. Time crystal behavior and S$\tau$B were observed in laser-driven Erbium samples~\cite{chen_realization_2023}. The effective Hamiltonian driving the system is time-independent, but oscillations of finite frequency were reported for the cross-correlation function between measured signal and the reference pulse, signaling a CTC. The system was analyzed under Lindblad dynamics, but is effectively few-body, as evidenced by the mean-field analysis. 

More recently, realizations of a CTC in an electron-nuclear spin system of a semiconductor have been reported~\cite{greilich_robust_2024}, where the system undergoes auto-oscillations for very long times. By employing magnetic fields, the electron spin polarization precesses through a process of dynamic nuclear polarization, resulting in observed collective CTC behavior, with observed lifetimes exceeding one hour. In Ref.~\cite{greilich_exploring_2025}, the same solid-state device was employed in the presence of a periodic drive achieved by periodically modulating the excitation polarization. This in turn results in the observation of collective oscillations that are $nT$-periodic, i.e., the system becomes a DTC. Collective synchronization between CTCs due to diffusion and mediated coupling processes has been recently reported in experiments using these solid-state devices~\cite{greilich_non-local_2025}. The above implementations employing semiconductors directly address the thermodynamic limit (with samples containing on the order of $10^{15}$ nuclear spins).


A dissipative DTC was observed~\cite{osullivan_signatures_2020} in low-temperature, microwave pulse-addressed doped $^{28}$Si. In the experiment, initially polarized spins are driven by a ``spin-locking" protocol that allows the spins to interact following an effective central spin model. The resulting order parameter, the spin polarization, displays subharmonic oscillations characteristic of DTC behavior in the presence of dissipation due to magnetic impurities and charge noise. DTCs stabilized by dissipation have also been reported in recent experiments involving atom-cavity platforms of BECs~\cite{kesler_observation_2021}, where the relative phase between the pump and the cavity light field undergoes subharmonic oscillations in moderate modulation regimes. Additionally, the intracavity photon number $N_P(t)$ undergoes S$\tau$B. This TC phase shows robustness against moderate white Gaussian noise levels.

Transitions between different time-crystalline regimes in dissipative environments have also been experimentally realized recently. In Ref.~\cite{cosme_torus_2025}, a BEC from $^{87}$Rb atoms was strongly coupled to an optical cavity, witnessing a transition from a dissipative CTC to a time quasi-crystal with two dominant subharmonic frequencies. The dissipative TC phase undergoes S$\tau$B in the cavity photon occupation, with observed lifetimes of a few milliseconds.

Dissipative time crystals have also been observed in Kerr-nonlinear optical microcavity setups described by effective classical theories. In Ref.~\cite{taheri_all-optical_2022}, a dissipative DTC was realized by driving the microresonator through injection locking by two independent laser pulses. The formation of periodic optical solitons around the resonator signals S$\tau$B in the system.

Strongly interacting Rydberg gases have also been employed in experiments to realize time crystal behavior, mostly comprising dissipative time crystals. A dissipative CTC was observed in Ref.~\cite{wu_dissipative_2024} employing a $^{85}$Rb vapor cell at room temperature traversed by parallel probe and reference beams, and a counter-propagating coupling beam that causes a subharmonic response due to coupling between levels. Multiple time crystals, including DTCs and high-harmonic time crystals, have also been realized by similar protocols involving Rydberg ensembles~\cite{jiao_observation_2025,jiao_photoionization-induced_2025,liu_bifurcation_2025}. In Ref.~\cite{smits_observation_2018}, a time-crystalline response was observed employing a superfluid quantum gas of ultracold atoms forming a BEC. In this case, the observation corresponds to a $2T$-DTC admitting an effective few-mode model description with almost negligible dissipation. In Ref.~\cite{liu_enhanced_2026}, a CTC in a dissipative Rydberg atomic gas was experimentally controlled near its critical boundary to a thermal phase, in order to achieve multi-parameter estimation in the context of quantum metrology.

Rydberg gases in the presence of an electric field have also been employed to witness time crystal behavior~\cite{arumugam_electric-field_2025,arumugam_stark-modulated_2025,arumugam_injection_2025}. In this case, strong electric fields are employed in order to obtain discrete time crystal behavior under the Stark effect. The system is dissipative, which can in principle prevent localization of the dynamics (rather resulting in a stable steady state limit after S$\tau$B).

Feedback schemes are currently being explored in the context of time crystals. Ref.~\cite{wang_observation_2025} reported the observation of time crystal behavior on a driven Rb-Xe spin-maser system with a feedback-induced retarded interaction, which causes S$\tau$B in the system if the feedback is strong enough. The order parameter is the magnetization of Rb atoms along the $x$ axis, with observed lifetimes reaching $80$s. The study concludes that the behavior of the time crystal in this case originates from the delayed feedback interaction.

Time crystals resulting from magnon dynamics in BECs have also been reported~\cite{kreil_tunable_2019,autti_ac_2021,autti_nonlinear_2022,trager_real-space_2021,makinen_time_2025}. The systems under consideration normally admit an effective classical theory that incorporates nonlinear effects, establishing a connection between the field of nonlinear wave propagation dynamics and Floquet time crystals.

A CTC from polaritons resulting from continuous wave optical excitations has been observed in Ref.~\cite{carraro-haddad_solid-state_2024}, where the photoluminescence spectra show the development of a subharmonic peak as the optical excitation power is increased. In this case, the dynamics of the reservoir are essential to observe S$\tau$B in the system.

Moving away from purely periodic temporal patterns, time quasicrystals were first observed in magnon condensation in superfluid $^3$He~\cite{autti_observation_2018} forming a BEC.

The experimental capabilities span different regimes of scale, control and stability. Atom–cavity and BEC realizations are tunable, allowing transitions between continuous, discrete and quasiperiodic time-crystalline responses to be resolved. Their observed lifetimes are typically limited to the millisecond range. Atomic and spin-gas platforms similarly permit controlled exploration of stability boundaries and noise resilience, while feedback schemes can extend the response to tens of seconds. The most persistent oscillations are found in macroscopic solid-state spin ensembles, where collective synchronization and nuclear-spin dynamics yield robust responses lasting up to hours, albeit with substantially less microscopic control over individual degrees of freedom. Compared with MBL and prethermal platforms, these systems can directly access the thermodynamic limit. They are less controllable and mainly provide access to global observables such as magnetization, making it difficult to establish genuine spatiotemporal order in many-body quantum correlations.

\subsection{Others / Unknown} 

\begin{figure}[!ht]
 \centering
        \includegraphics[scale=0.28]{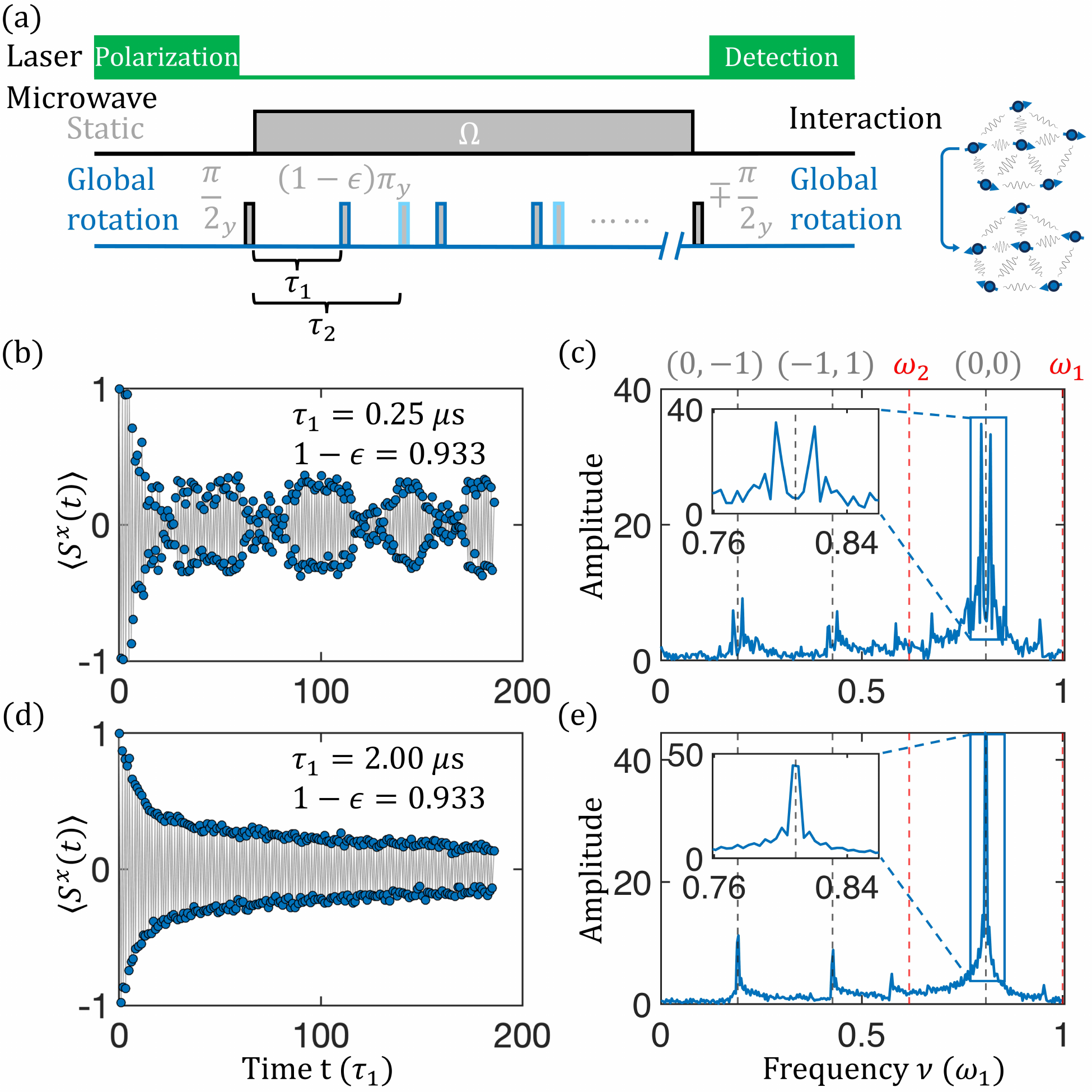}
 \caption{(Obtained from Ref.~\cite{he_experimental_2025}) Observation of a discrete-time quasicrystal. In Fourier space, discrete time quasicrystals are characterized by the emergence of several subharmonic contributions distinct from the two incommensurate frequencies $\omega_1$, $\omega_2$. This results in observing different quasi-periodic temporal patterns in the transverse spin polarization $\langle S^x(t)\rangle$. }
\label{fig5}
\end{figure}

By now, a variety of time crystals have been realized experimentally, showing stabilization mechanisms away from the three conventional ones previously discussed. Beyond other and unknown mechanisms, we also include here cases which cut across the MBL/prethermal/ECT distinction or combine several ingredients. 

A remarkable class of DTCs has been observed in disordered dipolar ensembles~\cite{choi_observation_2017}, corresponding to the so-called \emph{critical time crystal} discussed in Sec.~\ref{subsec:closed_sys}. In the experiment, an ensemble of NV impurities in diamond is used to encode a system of spins with $\sim J_{ij}r_{ij}^{-3}$ interactions, placing the system at the marginal case $\alpha = d$, where localization and delocalization compete. Under periodic driving, S$\tau$B is observed in the spin polarization signaling the emergence of DTC behavior. A similar experiment involving nitrogen-vacancy centers was carried out in Ref.~\cite{choi_probing_2019}, where thermalization dynamics of DTCs involving $\mathbb{Z}_2$ and $\mathbb{Z}_3$ symmetries were probed. In this case, the relevant order parameter experiencing S$\tau$B is the spin polarization of the ensemble, with observed lifetimes of $5.2\,\mu$s.

DTCs have also been observed in ordered dipolar spin ensembles~\cite{rovny_observation_2018,rovny_p_2018}, where nuclear magnetic resonance (NMR) data was reported to feature subharmonic response in spatiotemporal correlations and the average magnetization. The highly ordered 3D structure of these systems rules out MBL as the underlying stabilization mechanism. Prethermalization is also ruled out as a mechanism due to the high-temperature (with respect to the effective Floquet Hamiltonian) initial states employed. In Ref.~\cite{pal_temporal_2018}, nuclear spin clusters in molecules described by central spin models displayed DTC behavior in the low-frequency regime of the drive, also ruling out prethermal stabilization.

Recently, in Ref.~\cite{bao2026fockspace} a DTC has been realized through Fock space prethermalization (FSP) for systems consisting of 72 superconducting qubits. The normalized mean Hamming distance experiences S$\tau$B starting from initially ordered states, and this measure is directly correlated with localized wavepacket dynamics in the many-body wavefunction of the system.

Recently, a $3T$-periodic DTC from a disordered Floquet chiral clock model has been realized employing superconducting transmon qudits~\cite{goss2026qutrittimecrystalstabilized}. For a 15-qutrit model, results report clear signatures of 3$T$-periodic subharmonic oscillations in the local autocorrelation functions. The presence of fine-tuned chiral interactions in combination with disorder leads to observing time-crystal behavior for a variety of initial states configurations, whereas in the absence of chirality only few initial states experience S$\tau$B, a behavior that is beyond conventional MBL-DTC stabilization. 


In Ref.~\cite{chen2026}, $4T$-periodic oscillations in the local magnetization dynamics that are robust against gate errors and initial-state variations for an 8-qubit model have been reported. This occurs in the absence of disorder, and it is argued that the stabilization mechanism does not rely on dissipation or high frequency drives.

Stabilized DTCs due to structured noise have been implemented on superconducting-qubit hardware for periodically-kicked Ising models with uniform interactions and fields in alternative lattice geometries like the Kagome or Lieb lattice for 82 and 40 qubits, respectively~\cite{shinjo_emergent_2025}. The discrete S$\tau$B is attributed to two different mechanisms in this case, namely the existence of protected boundary modes or noise-induced period-doubling oscillations in local autocorrelation functions and total magnetization. Autocorrelation functions at different locations of the graph show distinct behavior, for example, in the Kagome lattice case, bulk correlations decay quickly towards a thermal phase behavior, while edge autocorrelation functions show robust period-doubling for longer times.

The experimental realization of time quasicrystals is a very promising emerging field with potential to unveil unknown stabilization mechanisms. For instance, recent realizations of discrete time quasicrystals have taken place in NV centers on $^{13}$C ordered structures~\cite{he_experimental_2025}, see Fig.~\ref{fig5}, where the stabilization mechanism is rather suspected to be intimately related to critical time crystals~\cite{choi_observation_2017}. Quasiperiodic S$\tau$B dynamics have also been observed employing superconducting quantum circuits~\cite{shinjo_unveiling_2024}. Experiments in strongly interacting Rydberg atoms have reported the observation of a time quasicrystal~\cite{Zhu2026}, where the probe light transmission spectrum develops subharmonic peaks at incommensurate frequencies.

Finally, we will mention that, despite being conceptually different from the many-body time crystal, implementations of photonic time crystals~\cite{wang_observation_2022,wang_metasurface-based_2023,xiong_observation_2025,liu_direct_2025} are currently being explored. Photonic time crystals undergo a subharmonic response in the refractive index at the single-particle level due to the system having a nontrivial topology. Including a nonlinear medium could potentially allow for interactions between the photons as in spontaneous parametric down-conversion (SPDC) processes. Photonic time crystals have been realized in experiments by applying mode-locked lasers in semiconductors, first proposed in Ref.~\cite{koch_pulse_2024} and recently experimentally realized~\cite{weng_time_2025}.

These realizations span different operational regimes. Dipolar spin ensembles provide platforms with long-range interactions, where neither conventional MBL nor high-frequency prethermalization readily explains their stability. Programmable quantum processors provide a flexible platform for exploring topological implementations of time crystals, where temporal order can appear at boundaries or in nonlocal observables, although the interplay between topology and other stabilization mechanisms remains to be fully understood. Qudit and temporally engineered platforms further extend the accessible phenomenology through tailored interactions and driving protocols, enabling higher-order and more general forms of temporal order. Thus, unlike MBL, prethermal or effective-classical realizations, no single mechanism determines the characteristic scale, lifetime or observable of these systems; instead, these experiments indicate possible routes for engineering where and how time-crystalline order emerges.

\subsection{Overview}\label{subsec:classifications}

A classification of time crystals is desirable to establish a connection between the various time crystals and the available quantum devices for their implementation. Until now, only a few classes of time crystals have been experimentally realized on quantum devices, but it is expected that new theoretical proposals will be implemented in the near term. Table~\ref{tab:table1} summarizes different types of time crystals, along with their ergodicity-breaking or stabilization mechanisms discussed in Secs.~\ref{subsec:closed_sys} and~\ref{subsec:open_sys}. 

The rows of Tab.~\ref{tab:table1} should be read as phenomenological characters of temporal order, while the columns indicate the dominant stabilization or ergodicity-breaking mechanism. These categories are not strictly mutually exclusive: a topological DTC may also be prethermal, a controlled TC may employ feedback on top of an otherwise MBL mechanism, and a critical TC may arise from the competition between disorder and long-range interactions. The table is therefore intended as an organizing map of present evidence and theoretical possibilities, rather than as a mutually exclusive taxonomy of phases. In particular, entries marked as observed may represent different levels of evidence, ranging from robust finite-time signatures on noisy devices to stronger indications of an asymptotic phase.


\begin{table}[h]
\centering
\caption*{\textbf{Overview of Time Crystals on Quantum Devices}} 
\begin{tabular}{@{} l *{5}{S[table-format=2.2]} @{}}
    \toprule
     & \multicolumn{4}{c@{}}{Ergodicity breaking / stabilization mechanism} \\
    \cmidrule(l){2-5}
    TC type  & {MBL} &  {Prethermal} & {ECT/few-body} & {Other/Unknown}   \\
    \midrule
    Closed DTC    &   \greencheck\textsuperscript{\footnotesize\cite{zhang_observation_2017,mi_time-crystalline_2022}} & \greencheck\textsuperscript{\footnotesize\cite{kyprianidis_observation_2021}} & \greencheck\textsuperscript{\footnotesize\cite{smits_observation_2018}} &  \greencheck\textsuperscript{\footnotesize\cite{rovny_observation_2018,pal_temporal_2018}}  \\
    Dissipative CTC    &   \orangeint & \orangeint & \greencheck\textsuperscript{\footnotesize\cite{kongkhambut_observation_2022}} &  \orangecheck\textsuperscript{\footnotesize\cite{russo_quantum_2025}}   \\ 
    Dissipative DTC    &   \orangeint & \orangecheck\textsuperscript{\footnotesize\cite{vu_dissipative_2023}} & \greencheck\textsuperscript{\footnotesize\cite{kesler_observation_2021}} & \greencheck\textsuperscript{\footnotesize\cite{shinjo_emergent_2025}}   \\ 
    Critical    &   \orangeint & \greencheck\textsuperscript{\footnotesize\cite{beatrez_critical_2023}}    & \orangeint & \greencheck\textsuperscript{\footnotesize\cite{choi_observation_2017}}   \\ 
    Controlled            &  \orangecheck\textsuperscript{\footnotesize\cite{camacho_prolonging_2024}}& \greencheck\textsuperscript{\footnotesize\cite{moon_discrete_2024}}  & \greencheck\textsuperscript{\footnotesize\cite{wang_observation_2025}} & \orangeint   \\ 
    Topological           &   \greencheck\textsuperscript{\footnotesize\cite{zhang_digital_2022}}  & \greencheck\textsuperscript{\footnotesize\cite{xiang_long-lived_2024}} & \orangeint & \orangeint   \\
    Time quasicrystals    &  \orangecheck\textsuperscript{\footnotesize\cite{zhao_floquet_2019}} & \orangecheck\textsuperscript{\footnotesize\cite{else_long-lived_2020}} & \greencheck\textsuperscript{\footnotesize\cite{autti_observation_2018}}  & \greencheck\textsuperscript{\footnotesize\cite{he_experimental_2025}}  \\  

    Time rondeau crystals    &   \orangecheck\textsuperscript{\footnotesize\cite{zhao_temporal_2023}} & \greencheck\textsuperscript{\footnotesize\cite{moon_experimental_2025}} & \orangeint  & \orangeint  \\
    \bottomrule
    \end{tabular}
    \caption{Table of observed \greencheck, unobserved \orangeint and theoretically predicted \orangecheck realizations of time crystals on quantum devices, depending on the ergodicity breaking or stabilization mechanism. We note that MBL and prethermal mechanisms are common routes to stabilize S$\tau$B in closed systems, whereas ECT or few-body descriptions are common in systems in the presence of dissipation. References are representative rather than exhaustive.} 
    \label{tab:table1}
\end{table}

The first category concerns systems realized in the ideal limit of an almost noise-free or error-corrected quantum device, where the time evolution in the system can be regarded to a good extent as unitary (\emph{closed systems}). As previously discussed in Sec.~\ref{subsec:origins}, if a discrete symmetry is broken and the system responds in correlation with the drive, i.e., subharmonically, this leads to the observation of a DTC~\cite{yao_time_2018}. To date, the observation of DTCs is limited by the coherence times of currently available noisy devices, but subsequent hardware improvement is expected to bring these time crystal realizations to firmer grounds.

\emph{Dissipative} time crystals comprise systems whose dynamics cannot be understood without taking into account effects from the surrounding environment. Since a certain degree of dissipation takes place between the system and environment, the evolution of the system is no longer unitary. The presence of dissipation allows the system to leak energy with the (macroscopic) environment, and thus the stabilization mechanism of these time crystals relies on the emergence of a steady state or a limit cycle that breaks any time-translational symmetries of the system. Dissipative DTCs and CTCs have been mainly implemented on platforms of BECs and Rydberg atoms. In particular, CTCs have until now not been observed in a (in principle) noise-free scenario.

A different category comprises those time crystals observed in disordered dipolar ensembles, also known as \emph{critical}. In critical time crystals, the decay of the dipolar interaction strength with distance is an essential feature competing against the presence of disorder in the system, as discussed in Sec.~\ref{subsec:closed_sys}. These types of time crystals have gained considerable relevance in the field of quantum simulation employing NV center devices.

Another type of time crystals concerns \emph{controlled} systems and systems hosting \emph{topological} properties. The reason to consider these two cases separately is the potential direct connection these TCs have with quantum error correction (QEC) protocols. Even though for quantum error correcting codes the whole system of logical and physical qubits can be considered as a closed system, the logical subspace is naturally an open system being entangled with the rest of the qubits necessary for error correction. Thus, these time crystals, if implemented, would resemble neither purely closed nor open systems.

There has been recent interest in moving away from the concept of a perfect crystal in time, relaxing the periodicity condition in order to allow for quasi-ordered spatiotemporal patterns in many-body systems. Thus, we also include the novel route to witness partial time order in driven systems, those known as time \emph{quasicrystals}~\cite{zhao_floquet_2019,dumitrescu_logarithmically_2018,else_long-lived_2020}. In analogy with conventional quasicrystals in space, time quasicrystals develop almost periodic patterns along the time dimension. More concretely, a discrete time quasicrystal is a system responding to a quasiperiodic drive with a set of frequencies that are \emph{not} harmonics of the original drive. If the original drive is quasiperiodic, it will generally have a spectral decomposition with peaks at frequencies $\omega_p = \sum_l n_{l}^{(p)}\omega_l^{(p)}$, i.e., linear combinations of some incommensurate frequencies. The discrete time quasicrystal breaks this apparent order by responding quasiperiodically with a different set of frequencies $\tilde{\omega}_p$ than those of the drive~\cite{zaletel_colloquium_2023}.

We also include new advances in systems where spatiotemporal order at stroboscopic times coexists with temporal disorder in the micromotion. The idea is that even though the Floquet theorem does not apply to these temporally disordered systems, prethermalization~\cite{mori_thermalization_2018} can still take place in the system, e.g., by employing random multipolar drives~\cite{zhao_random_2021,mori_rigorous_2021,zhao_temporal_2023}. Even though spatiotemporal correlations decay algebraically in this case, employing large dipolar structured drives can significantly increase the lifetimes of the prethermal phase, even to exponential scale. These new systems realized through non-periodic, but otherwise structured drives, have been called \emph{time rondeau crystals}.

\begin{table}[t]
\centering
\footnotesize
\begin{threeparttable}
\caption*{\textbf{Comparative table for time crystal experiments}}
\begin{tabular}{@{} l *{8}{c} @{}}
\toprule
Work & Platform & Type & Stabilization & Size & Lifetimes & Observable & Interactions & Cycles \\
\midrule

\cite{mi_time-crystalline_2022} & SC & DTC & MBL & 20 & $\sim 5\mu$s & $\langle\overline{\sigma^z(nT)\sigma^z(0)}\rangle$ & NN & 60\\

\cite{zhang_observation_2017} & TI & DTC & MBL(*) & 10 & $\sim 7$ms & $\langle\sigma^x(nT)\sigma^x(0)\rangle$ & $r_{ij}^{-1.5}(<1\text{kHz})$ & 100\\

\cite{frey_realization_2022} & SC & DTC & MBL & 57 & $\sim 40\mu$s & $\langle\overline{\sigma^z(nT)\sigma^z(0)}\rangle$ & NN & 50\\

\cite{randall_many-bodylocalized_2021} & NV & DTC & MBL & 9 & $\sim 0.5$s & $\langle\overline{\sigma^z(nT)\sigma^z(0)}\rangle$ & $r^{-3}(\sim 1-10$ Hz) & $\sim$800\\

\cite{switzer_realization_2025} &SC & DTC & MBL & 156 & $\sim 40\mu$s & HD,QFI & NN & $\sim$40\\

\cite{zhang_digital_2022} & SC & Topological & MBL & 26 & $\sim 10\mu$s& $\langle Z_{L_{i}}(nT)\rangle$ & NN  & $\sim$40 \\

\cite{kyprianidis_observation_2021}& TI & DTC & PT & $25$ & $\sim 15$ms & $\langle\sigma^x(nT)\sigma^x(0)\rangle$ & $r_{ij}^{-\alpha}$($<1\text{kHz}$)& $\sim$50\\

\cite{moon_discrete_2024}& NV & Controlled & PT & $\sim 10^4$ & $\sim 10$s & $\langle P(nT)\rangle$ & $r_{ij}^{-3}$($\sim 0.6\text{kHz}$)& $\sim$$10^4$\\

\cite{xiang_long-lived_2024} & SC & Topological & PT & 18 & $\sim \mu$s& $\langle Z_{L_{i}}(nT)\rangle$ & Stabilizer& 20 \\

\cite{moon_experimental_2025} & NV & TRC & PT & $\sim 10^{19}$ & $\sim 4$s& DFT$(\langle S(nT)\rangle)$ & $r_{ij}^{-3}(0.66$kHz) & $\sim$100 \\

\cite{beatrez_critical_2023} & NV    & Critical & PT & $\sim 10^4$ & $\sim 4.6$s &$\langle P(nT)\rangle$ & $r_{ij}^{-3}(\sim 0.66$kHz) & 396\\ 

\cite{kongkhambut_observation_2022} & BEC    & CTC & ECT/FB & $\sim 10^4$ & $\sim$ 2ms &$\langle N_P(t)\rangle$ & $\sim$kHz & $\sim$40\\  

\cite{smits_observation_2018} & SFQG    & DTC & ECT/FB & $\sim 10^7$ & $\sim$ 500ms & LD & $\sim 0.4$kHz & $\sim$50\\  

\cite{greilich_robust_2024} & SS    & CTC & ECT/FB & $\sim 10^{6}$ & $\sim$ 1\text{h} &FR & $\sim$kHz & $\sim$$10^3$\\

\cite{kesler_observation_2021} & BEC    & DTC & ECT/FB & $\sim 10^3$ & $\sim$ 2ms &$\langle N_P(t)\rangle$ & $\sim$kHz & $<$10\\ 

\cite{wang_observation_2025} & SMS    & Controlled & ECT/FB & $\sim 10^{17}$ & $\sim$ 80s &$\langle M(nT)\rangle$ & $\sim$ 30-90 Hz & $>$$10^3 $\\ 

\cite{shinjo_emergent_2025} & SC  & Diss.DTC & Unknown & 156 & $\sim 12\mu$s &$\langle \sigma^z(nT)\rangle$ & Graph-based & 40\\

\cite{autti_observation_2018} & BEC & DTQC & ECT/FB & $\sim 10^{12}$ & $\sim 20$s& $\langle M(nT)\rangle$ & $\sim$kHz & $\sim$200 \\

\cite{rovny_observation_2018} & ODE & DTC & Other & $\sim 10^{23}$ & $\sim 1.6$ms& $\langle P(nT)\rangle$ & $r_{ij}^{-3}(\sim 20$kHz)  & $\sim$80 \\

\cite{bao2026fockspace}& SC & DTC & Other & $72$ & $\sim 0.8\mu$s & HD& NN & $\sim$80\\

\cite{pal_temporal_2018} & NSC & DTC & Other & $37$ & $\sim 12$s& $\langle M(nT)\rangle$ & $\sim10$Hz  & $\sim$300 \\

\cite{he_experimental_2025} & NV & DTQC & Other & $\sim 10^4$ & $\sim 0.4$ms& $\langle M(nT)\rangle$ & $r_{ij}^{-3}(0.3$MHz)  & $\sim$200 \\

\cite{choi_observation_2017} & NV  & Critical & Other & $>10^3$ & $\sim 10$ms & $\langle P(nT)\rangle$ & $r_{ij}^{-3}(\sim 100$kHz) & $\sim$30\\

\bottomrule
\end{tabular}

\begin{tablenotes}
\footnotesize
\item \textbf{Abbreviations:} Diss.DTC=Dissipative DTC; DTQC=Discrete time quasicrystal; TRC=Time rondeau crystal; TI= Trapped-ion; SC = Superconducting; NV= Nitrogen Vacancy centers; NSC=Nuclear spin clusters; SS = Solid State; PT=Prethermal; BEC=Bose-Einstein Condensate; SFQG=Superfluid quantum gas; SMS=Spin-Maser system; ODE=Ordered dipolar ensemble; ECT/FB=Effective Classical Theory/Few-Body; NN=Nearest-neighbour; HD=Hamming distance; LD=Line density; QFI=Quantum Fisher information, FR=Faraday rotation
\end{tablenotes}
\caption{Comparison between implementations across different quantum platforms. The size refers to the total number of qubits employed in the simulation, whereas lifetimes refer to the approximate coherence time of the time crystal. The number of cycles refers to the total number of periods observed. The observables $\langle P(nT)\rangle$, $\langle N_P(t)\rangle$ and $\langle Z_{L_i}(nT)\rangle$ refer to the polarization, the intracavity photon number and non-local string operators, respectively. DFT$(\langle S(nT)\rangle)$ corresponds to the discret Fourier transform of a discrete signal $S(nT)$ at stroboscopic times in the time rondeau case. The reported observable has been chosen by significance.(*) See Refs.~\cite{khemani_comment_2021} and~\cite{yao2021replycommentdiscretetime} for a discussion on the stabilization mechanism.}\label{tab:table2}
\end{threeparttable}
\end{table}

We would like to stress that the classification framework presented here is not meant to be hierarchical, in the sense that it is not uncommon that different types of time crystals find overlap with more than one category. For instance, the so-called critical time crystals~\cite{choi_observation_2017,ho_critical_2017} are good candidates to share aspects of time crystals whose stabilization mechanism is driven both by MBL or prethermal behavior. The reason is that these systems present some degree of disorder, leaving the question open as to whether there could exist regimes where MBL survives in the presence of long-range interactions. On the other hand, prethermal stabilization has been demonstrated to exist in these systems~\cite{beatrez_critical_2023}. Thus, the question remains: Is it possible to have a critical time crystal stabilized by MBL or prethermalization, and if so, how is the crossover between these two regimes? Another candidate where alternative stabilization mechanisms might be involved is in controlled time crystals (see Table~\ref{tab:table1}). A likely situation in these systems is that continuous monitoring will partially dissolve the underlying stabilization mechanism (for instance, MBL), in favor of transitioning towards an ECT/few-body description, essentially due to the repeated non-unitary operations (measurements) taking place in the system. A detailed exploration of the interplay between these effects in controlled TCs could lead to further refinement in the classification of these monitored systems. In the context of open quantum systems, the interplay between high-frequency regimes and the coupling to the environment~\cite{vu_dissipative_2023} can lead to transition from a regime with prethermal stabilization to ECT/few-body due to strong dissipation effects.

The overview of the time crystal implementations in Tab.~\ref{tab:table1}  provides a qualitative description of most experiments carried out on a variety of quantum platforms. While this provides a rather pedagogical view on the status of quantum time crystals, it is important to highlight more quantitative aspects of the field, such as the size of the systems, the observed lifetimes and accessible observables on the different platforms.
We summarize these aspects in Tab.~\ref{tab:table2} for a representative class of platforms, DTC type and stabilization mechanisms. No single platform simultaneously maximizes system size, lifetime and microscopic control, highlighting the complementary capabilities of present time-crystal implementations.

We comment on the evidence of quantum correlations reported by TC experiments. To date most of the works on TC implementations have focused on addressing relevant order parameter dynamics such as total magnetization or local autocorrelation functions to witness S$\tau$B occurring in the system. While this might be sufficient to observe the subharmonic periodic motion, addressing genuine many-body correlations offers a more valuable exploratory ground to understand the quantum nature of TC behavior. In this regard, it is very desirable that future implementations aim to further observe varied spatio-temporal correlations, as well as more information-theoretic quantities directly related to entanglement in these systems. We believe that obtaining experimental evidence for these quantities provides promising potential for developing a systematic classification of TCs based on their intrinsic quantumness.

\section{Further proposals}\label{sec:proposals}

Over the past decade, research on time crystals has generated a broad range of potential applications, with theoretical developments advancing rapidly~\cite{zaletel_colloquium_2023,hannaford_decade_2022}. Here we outline several active research directions focused on proposals for realizing quantum time crystals that may be implementable on near-term quantum devices. We note that these proposals are included as possible future realizations but have not yet reached the conceptual maturity of the works discussed above, where implementations have already been carried out. Given the vast literature on the theoretical proposals, we refrain from giving a systematic review, but rather limit ourselves to a broad overview focusing only on the proposed stabilization mechanism and time crystal type.

The most promising proposals for immediate implementation on current quantum hardware concern time crystals in the context of clean (disorder-free) DTCs. Some proposals to realize clean DTCs involve periodically driven quantum spin chain models~\cite{yu_discrete_2019,medenjak_isolated_2020,yarloo_homogeneous_2020,liang_time_2020,choudhury_route_2021,nurwantoro_discrete_2019,bar_lev_discrete_2024}, two-dimensional DTCs~\cite{santini_clean_2022}, kicked boson~\cite{giergiel_discrete_2023} and hardcore-boson models~\cite{yang_dynamical_2021}, clock models hosting long-range interactions~\cite{surace_floquet_2019} and models of Rydberg atoms~\cite{fan_discrete_2020,xue_enhanced_2026}. Most of these fall under the Other and ECT/few-body categories.
Periodically driven spin-ladder models employing ultra-cold atoms to realize time crystal behavior have been put forward~\cite{huang_clean_2018}. In this case, the couplings among different directions of the spin ladder are alternated within a single period of the drive. The stabilization mechanism for these time crystals is shown to go beyond the prethermal regime even in the presence of strong interactions. Another route to construct clean DTC models relies on the presence of Floquet dynamical symmetries~\cite{chinzei_time_2020,medenjak2020}, gauge symmetries~\cite{russomanno_homogeneous_2020} or integrability~\cite{PhysRevB.100.184301}.

A widely studied approach to clean TCs involves the application of high-frequency driving, which induces a prethermal regime in which S$\tau$B may occur. These approaches generally rely on high-frequency expansions of the effective Floquet Hamiltonian~\cite{mizuta_high-frequency_2019}. Recent work has shown the robustness of this phase against transverse and longitudinal perturbative fields~\cite{fernandes_nonperturbative_2025}, with some prethermal phases displaying critical properties~\cite{natsheh_critical_2021,natsheh_critical_2021-1}. It has also been proposed that Hilbert space fragmentation can lead to very long-lived DTCs in the absence of disorder even at moderately small values of the driving frequency~\cite{tang_discrete_2025}. Even though prethermal DTCs are in general strongly dependent on the initial state configuration, it has been recently proposed that prethermalization can exist for arbitrary initial states in models hosting zero-dimensional corner modes~\cite{jiang_prethermal_2024}. There have been several proposals to realize disorder-free time crystals in quantum dot systems~\cite{sarkar_emergence_2022,sarkar_time_2024}.

For disordered systems undergoing MBL, the MBL-DTC for the kicked, disordered Ising chain and its variants has now been studied extensively, remaining an active area of research as a paradigmatic model for DTC behavior. Recent results show the connection between temporal spin correlations and spectral characteristics of the model~\cite{Schmid2024,penner_subharmonic_2025}. It is known that MBL is the driving ergodicity-breaking mechanism; however, understanding the phase diagram in full generality remains an active area of current research~\cite{zhang_subexponential_2023,liu2025analytical}. Away from the standard MBL mechanism based on short-ranged interactions, alternative localization mechanisms based on the Stark effect involving strong electric fields have been proposed to observe time crystal behavior~\cite{kshetrimayum_quantum_2021,kshetrimayum_stark_2020,liu_discrete_2023,wang_discrete_2025}. Other proposals involve chiral Floquet phases of interacting bosons in combination with MBL~\cite{po_chiral_2016}. The combination of disorder and prethermal DTCs at very high frequencies has been explored in quantum spin chain models~\cite{zeng_prethermal_2017}. The simulation of two-dimensional models~\cite{sims_simulation_2023} and systems forming highly geometric patterns~\cite{sims_formation_2023} in the presence of disorder further support the possibility of implementing higher-dimensional MBL-DTCs employing superconducting qubit hardware, as well as quantum dots~\cite{foulk_realizable_2023,throckmorton_effects_2022} and diamond-based~\cite{egawa_controlling_2025} quantum simulators.

Topological quantum time crystals~\cite{giergiel_topological_2019} have also been proposed, with only a few implementations realized so far on quantum devices, as discussed in Sec.~\ref{sec:implementations}. Some of these proposals involve the presence of interactions in systems undergoing MBL~\cite{wahl_topologically_2024,motamarri_symtft_2023}, magnon dynamics in interacting spin systems~\cite{bhowmick_discrete_2023} as well as tight-binding models subjected to periodic drive~\cite{peng_topological_2022,wang_topological_2022}. The unification of topological quantum systems and nonequilibrium dynamics shows great potential in finding genuine time crystal phases~\cite{placke_topological_2024} in cases where the stabilization mechanism is rather unknown (see Table~\ref{tab:table1}).

In the context of dissipative time crystals, several proposals have gathered considerable attention. One of them is boundary time crystals~\cite{iemini_boundary_2018,piccitto_symmetries_2021,prazeres_boundary_2021,nakanishi_dissipative_2023,xu_boundary_2023,carollo_quantum_2024,jirasek_boundary_2025,Paulino_2026}, where a small portion of the Hilbert space undergoes S$\tau$B, whereas the rest of the system remains time-translationally invariant and is regarded as an environment. Due to the low dimensionality of the system, some boundary time crystal models admit exact solutions~\cite{carollo_exact_2022}. In addition, these systems can host multipartite entanglement in the steady state~\cite{lourenco_genuine_2022, montenegro_quantum_2023}, and are expected to be highly relevant in the development of quantum sensing applications~\cite{iemini_floquet_2024,cabot_continuous_2024,yousefjani_discrete_2025,gribben_boundary_2025,shukla_prethermal_2025,yousefjani_prb_2025,yousefjani2026nonlinearityenhancedquantumsensingdiscrete}. Recent works have explored the role of long-range interactions in boundary time crystals~\cite{wang_boundary_2025}, the presence of dynamical quantum phase transitions~\cite{mondkar2026dynamical} and topological properties~\cite{nemeth2026topological,nemeth2026operatorspacetransportemergence}.


Early proposals to realize time crystals in atom-cavity setups of BECs~\cite{kesler_emergent_2019,kesler_continuous_2020} have led to an immense activity in recent years involving dissipative time crystals. These include the survival of the DTC phase in the presence of dissipation for one-dimensional disordered~\cite{lazarides_fate_2017} and ordered spin systems~\cite{lazarides_time_2020}, fermionic~\cite{buca_non-stationary_2019,booker_non-stationarity_2020}, bosonic~\cite{lledo_dissipative_2020,esencan2026timecrystalspassivelyprotected} and spin-boson models~\cite{souza_sufficient_2023}, as well as spin systems with long-range interactions~\cite{tucker_shattered_2018,chinzei_criticality_2022,wang_time-crystal_2023,passarelli_dissipative_2022,cabot_quantum_2023} or central spin models~\cite{cabot_metastable_2022}. Dissipative time crystals have also been predicted to exist for quantized charge density wave (CDW) rings coupled to an environment, where a metastable ground state of the system is formed and displays periodic oscillations~\cite{nakatsugawa_quantum_2017}. In this case, decoherence acts as the source of S$\tau$B. Realizing limit cycles resembling time crystal behavior in dissipative systems has also been explored in two-dimensional spin models~\cite{yang_emergent_2025} as well as chiral atomic systems evolving under nonlinear master equations~\cite{silveirinha_spontaneous_2023}. These cases can be regarded as a subclass of CTC emerging through dissipation or the presence of self-organizing, non-Markovian dynamics depending on the initial state choice~\cite{xiang_self-organized_2023}. Complementing these findings, it has been reported that the dynamics of CTC in the presence of dissipation cannot be faithfully captured by mean-field approaches if the initial state of the system is highly correlated~\cite{mukherjee_symmetries_2024}. The nonequilibrium transition occurring in dissipative DTC models has also been explored in the periodically driven quantum van der Pol model~\cite{cabot_nonequilibrium_2024}. More recently, chaotic synchronization~\cite{postavova_classical_2025} and the emergence of continuous quasicrystal order~\cite{solanki2025} in coupled dissipative time crystals have been explored, as well as models involving non-Markovian dissipation~\cite{das_discrete_2024} that are particularly relevant for systems composed of interacting Rydberg gases~\cite{gambetta_discrete_2019}.


Periodically driven BECs are regarded as natural setups to realize dissipative time crystal behavior. It has been shown that BECs coupled to a cavity can display time crystal behavior in the thermodynamic limit~\cite{buca_dissipation_2019} or lead to a metastable DTC phase~\cite{tuquero_dissipative_2022}, with recent proposals for few-mode models~\cite{cosme2026timecrystalscavitybecsystems}. Along these lines, numerous proposals involving cavity QED and Kerr resonators to realize time crystals in the presence of dissipation have been put forward~\cite{cole_subharmonic_2019,cosme_time_2019,gong_discrete_2018,zhu_dicke_2019,jara_theory_2024,mattes_entangled_2023,jager_dissipative_2024,chen_discrete_2024}, even in the presence of nonlinear effects~\cite{li_time_2024}.

An interesting avenue is to realize time crystals in ultra-cold atom ensembles bouncing on a periodically oscillating mirror~\cite{giergiel_time_2018,kuros_phase_2020,giergiel_creating_2020,hannaford_condensed_2022,golletz_basis_2022,golletz_formation_2025}. In this case, the atoms strongly attract each other, forming a BEC, and the overall dynamics can be effectively described by a Gross-Pitaevskii equation and mean-field approximation. Further theoretical approaches in the context of BECs to realize time crystals have been put forward~\cite{liao_dynamics_2019,lledo_driven_2019,ohberg_quantum_2019,wang_many-body_2021,wang_discrete_2021,martins_polaritonic_2022,johansen_role_2023}. This can have important implications for synchronization approaches in CTCs realized in BECs~\cite{solanki_exotic_2024}. Proposals for realizing emergent limit cycles on atom-cavity systems also exist~\cite{kesler_emergent_2019}.



In the area of quantum computation and quantum information processing, there have been a number of proposals to develop quantum error correction using time crystals~\cite{bomantara_quantum_2021,bomantara_nonlocal_2021,bomantara_simulation_2018,bomantara_square-root_2022}. This subfield of research comprises the controlled time crystals shown in Table~\ref{tab:table1}. Systems defined in the context of monitored or measurement-based quantum systems deserve special mention in this category~\cite{krishna_measurement-induced_2023,mcginley_absolutely_2022,oconnor_quantum-enhanced_2025}, particularly concerning monitored quantum clocks~\cite{viotti_quantum_2025}. Recent approaches explore the application of measurement-based feedback schemes to time crystal stabilization in the presence of decoherence~\cite{camacho_prolonging_2024,cenedese_thermodynamics_2025}. 

It is possible to relax the strict periodic dynamical constraint of time crystals by applying quasiperiodic structures in time. These systems include time quasicrystals~\cite{dumitrescu_logarithmically_2018,else_long-lived_2020,zhao_floquet_2019,pizzi_period-_2019}, including in dissipative environments~\cite{feng_tunable_2025}. More exotic systems include engineering time crystals in the time domain~\cite{giergiel_inseparable_2021,giergiel_discrete_2019} and timetronics~\cite{giergiel_time-tronics_2024}. Going away from systems with pure periodicity, but with otherwise structured drives in time domain, temporal disorder in the micromotion coexists with long-range order at stroboscopic times: these are the so-called \emph{time rondeau crystals}. In this case, high-frequency driving similar to the case of prethermal DTCs is the leading stabilization mechanism for initial states breaking the discrete symmetry of an effective Hamiltonian; nevertheless, correlations are expected to decay algebraically~\cite{zhao_temporal_2023}. The stability of time rondeau crystals in the presence of dissipation is a current topic of interest~\cite{ma_stable_2025}.

Many theoretical proposals have explored new platforms to realize time crystals. These include the realization of time crystals using superconducting Higgs modes~\cite{homann_higgs_2020,ojeda_collado_emergent_2021,dai_photo-induced_2022,fan_emergence_2024,kaplan2026}, phonon vibrational modes~\cite{zhan_time_2025}, hybrid Josephson junctions~\cite{nashaat_self-generated_2025}, time crystals emerging from excited eigenstates and many-body scars~\cite{huang_discrete_2022,syrwid_time_2017,bull_tuning_2022} or in few-body systems~\cite{barfknecht_realizing_2019}, and proposals to realize DTCs in long-range interacting systems~\cite{kozin_quantum_2019,lyu_eternal_2020,munoz-arias_floquet_2022,pizzi_bistability_2021,pizzi_higher-order_2021,russomanno_floquet_2017,cosme_bridging_2023,gargiulo_swapping_2024,reimann_nonequilibration_2023,rahaman_time_2024}.


In the quest to realize higher-dimensional time-crystalline phases using quantum hardware, proposals involving qudit platforms have also received considerable attention lately~\cite{ma_qudit-native_2025}. The possibility to observe time-crystalline behavior employing these devices~\cite{camacho_observing_2024} paves the way towards simulating these systems beyond qubit architectures, for which several theoretical works have been put forward~\cite{kshetrimayum_quantum_2021,kuros_controlled_2021,dziarmaga_simulation_2022,prokofev_algebraic_2018,giachetti_fractal_2023,chandra_discrete_2024,hu_solvable_2023}. More exotic proposals involve fractional and non-Hermitian DTCs~\cite{matus_fractional_2019,yousefjani_discrete_2025}, time crystals emerging from spatial translation-induced S$\tau$B~\cite{mizuta_spatial-translation-induced_2018}, coupled models of continuous and discrete time crystals~\cite{schumann_hierarchical_2026}, and recent theoretical developments in the study of out-of-time ordered correlators (OTOCs) crystals~\cite{buca_out--time-ordered_2022,hajdusek_seeding_2022}. 

Even though current applications involving DTCs are limited, there is ongoing interest in exploring the role time crystals can play in the context of quantum spin networks related to machine learning models and quantum reservoir computing~\cite{sakurai_dephasing-induced_2021,estarellas_simulating_2020,sakurai_chimera_2021,sakurai_quantum_2022,zhang_robust_2025}.

Finally, proposals for photonic devices that involve the realization of time crystals are varied~\cite{sharabi_disordered_2021,dikopoltsev_light_2022,lustig_topological_2018,saha_photonic_2023,wang_one-dimensional_2021,jin_floquet_2022,hayran__2022,lyubarov_amplified_2022} and are being actively explored in the context of metamaterials~\cite{asgari_theory_2024,dong_nonuniform_2025,li_stationary_2023,sadhukhan_defect_2023,sadhukhan_momentum_2023} and nonlinear photonic time crystals hosting superluminal k-gap solitons~\cite{pan_superluminal_2023}. Several works consider that there is an ECT mechanism behind these systems~\cite{kiselev_symmetry_2025}, which is usually the case in the presence of dissipation~\cite{seibold_dissipative_2020}. An interesting research route along these lines is to realize pairs of entangled plasmons in these systems~\cite{kiselev_light-controlled_2024}. Photonic time crystals have been proposed as systems in which Anderson's localization can be generalized to the time domain~\cite{eswaran_anderson_2025}.

\section{Conclusion}\label{sec:conclusion}

Based on our review, we find that various new types of time crystals, beyond the DTC and CTC dichotomy, have emerged and have already been realized experimentally. This demonstrates the capabilities brought by current quantum devices to study such nonequilibrium phases of matter and explore their properties. Nevertheless, a unifying framework covering all of these cases is still lacking, with several fundamental questions remaining open.

A particularly interesting direction concerns understanding the transitions between the variety of existing time crystal phases having different stabilization mechanisms as summarized in Table~\ref{tab:table1}. It is also relevant to understand whether such (necessarily nonequilibrium) transitions between time crystal phases occur abruptly, perhaps sharing some analogies with the standard theory of phase transitions in equilibrium, or whether they instead constitute a nonequilibrium crossover. 

Along these lines, exploring the effect of these transitions on entanglement and also stabilizer entropy~\cite{Leone2022} is a promising route to understand how quantum information changes between time crystal phases. Intuitively, the amount of entanglement existing in deep time crystal phases is expected to be moderate, considering that the system presents a rather close-to-classical dynamical behavior. By appropriate parameter tuning, the system can be brought to a critical regime separating two different time-crystal phases. In this regime, the existence of nonlocal correlations is expected to be enhanced and captured by several entanglement witnesses. Achieving this degree of control in time crystal dynamics has promising potential implications in the development of optimal strategies to manipulate and further exploit entanglement as a resource on different quantum devices. 

The experimental characterization of time crystals has progressed substantially beyond only observing a subharmonic response. In several implementations, time-crystalline order has been evaluated through its rigidity against perturbations, persistence over extended parameter regimes, dependence on initial conditions, lifetime and heating dynamics, and scaling with system size, while some experiments have additionally accessed spatial and spatiotemporal correlations. These measurements provide increasingly stringent operational corroboration for S$\tau$B and for the mechanisms stabilizing the observed phases. What remains comparatively unexplored, however, is the structure of the non-classical correlations accompanying this order. Only a limited number of experiments have directly addressed quantities sensitive to entanglement, higher order correlations and non classical measures of correlations based on quantum coherence. At the present level of experimental control, extending the characterization of time crystals to entanglement witnesses such as the QFI, as well as probes of spatially extended correlations, connected $n>2$ point functions and information scrambling such as OTOCs, therefore constitutes an important next step. Such measurements would complement the operational identification of time-crystalline order by revealing to what extent its stabilization and dynamical properties are intrinsically many-body, and could provide a basis for comparing the quantum character of time crystals across different phases and experimental platforms.

One important application of time crystals concerns the storage of quantum information in systems where interactions are otherwise expected to induce nontrivial dynamics and information scrambling. Time crystals overcome information loss through coherent superposition effects, making it possible to store various types of quantum data ranging from low-energy states, as in the prethermal TC, to general product states in the MBL-DTC case. This is particularly relevant in settings where the system is externally driven by its environment; 
in this case, appropriately engineering the interactions to realize a TC can prevent otherwise rapid thermalization (see Sec.~\ref{subsec:ergodicity}). Another relevant scenario concerns the notion of designing quantum memories for practical purposes through quantum low-density parity-check (qLDPC) codes~\cite{placke_topological_2024}. Elevating passive qLDPC codes to driven quantum memories is an interesting route in this domain.

Another promising route in this direction is that of stabilization mechanisms including non-unitary operations, i.e., measurement and feedback schemes, which can conveniently be incorporated during circuit execution times.

Time crystals have various potential applications within quantum technologies, particularly in quantum sensing and metrology~\cite{choi2017quantummetrologybasedstrongly}. As discussed in Sec.~\ref{sec:proposals}, many proposals consider time crystals as good candidates to develop highly reliable quantum sensors capturing disturbances in the subharmonic motion of the system due to external perturbations. This high sensitivity of time crystals to external disturbances can also be employed for quantum metrology tasks. For example, high-precision parameter estimation can be achieved by tuning a time crystal into a phase where the quantum Fisher information becomes significantly enhanced. 

Finally, the overview presented in Sec.~\ref{subsec:classifications} shows a promising window for near-term implementations that have yet to be explored, particularly those already theoretically predicted in Table~\ref{tab:table1}.

\acknowledgements{This work was funded by the Deutsche
Forschungsgemeinschaft (DFG, German Research Foundation)-Project No. FA 1884/5-1. Q-Neko project has received funding from the European Union’s Horizon Europe research and innovation programme under Grant Agreement No. 101241875. This work was also
performed for Council for Science, Technology and Innovation (CSTI), Cross-ministerial Strategic Innovation Promotion Program (SIP), “Promoting the application
of advanced quantum technology platforms to social issues”(Funding agency: QST). This project was made possible by the DLR Quantum Computing Initiative and the Federal Ministry for Research, Technology and Space of Germany.}

\bibliographystyle{apsrev4-2}
\bibliography{bib.bib}

\end{document}